\documentclass[%
 aps,
 amsmath,amssymb,
 reprint,%
]{revtex4-2}
\usepackage{graphicx}% Include figure files
\usepackage{dcolumn}% Align table columns on decimal point
\usepackage{bm}% bold math
\usepackage{color}
\usepackage{amsfonts}
\usepackage{mathtools}
\usepackage[utf8]{inputenc}
\usepackage[T1]{fontenc}
\usepackage{mathptmx}
\usepackage{etoolbox}
\usepackage{sidecap}

\makeatletter
\def\@email#1#2{%
 \endgroup
 \patchcmd{\titleblock@produce}
  {\frontmatter@RRAPformat}
  {\frontmatter@RRAPformat{\produce@RRAP{*#1\href{mailto:#2}{#2}}}\frontmatter@RRAPformat}
  {}{}
}%
\makeatother
\begin{document}

\preprint{AIP/123-QED}

\title[Dynamic Information Transfer in Stochastic Biochemical Networks]{Dynamic Information Transfer in Stochastic Biochemical Networks}
\author{Anne-Lena Moor$^{1,2}$}
\author{Christoph Zechner$^{1,2,3}$}%
 \email{zechner@mpi-cbg.de}
\affiliation{ 
$^1$Max Planck Institute of Molecular Cell Biology and Genetics, 01307 Dresden, Germany%\\This line break forced with \textbackslash\textbackslash
}%
\affiliation{ 
$^2$Center for Systems Biology Dresden, 01307 Dresden, Germany%\\This line break forced with \textbackslash\textbackslash
}%
\affiliation{$^3$Cluster of Excellence Physics of Life, TU Dresden, 01062 Dresden, Germany}

\date{\today}

\begin{abstract}
We develop numerical and analytical approaches to calculate mutual information between complete paths of two molecular components embedded into a larger reaction network. In particular, we focus on a continuous-time Markov chain formalism, frequently used to describe intracellular processes involving lowly abundant molecular species. Previously, we have shown how the path mutual information can be calculated for such systems when two molecular components interact directly with one another with no intermediate molecular components being present. In this work, we generalize this approach to biochemical networks involving an arbitrary number of molecular components.  We present an efficient Monte Carlo method as well as an analytical approximation to calculate the path mutual information and show how it can be decomposed into a pair of transfer entropies that capture the causal flow of information between two network components. We apply our methodology to study information transfer in a simple three-node feedforward network, as well as a more complex positive feedback system that switches stochastically between two metastable modes. 
\end{abstract}

\maketitle

\paragraph*{Introduction.} Information theory provides a powerful mathematical framework to study information transfer in complex dynamical systems. Originating from man-made communication systems \cite{shannon1948mathematical}, information theory has made its way into the biological sciences where it has helped to understand diverse biological processes, ranging from tissue patterning \cite{dubuis2013positional} to signal transduction \cite{Selimkhanov1370}.

A central concept in information theory is that of \textit{mutual information}, a quantity that captures the amount of information that is shared among two random objects $X$ and $Y$ \cite{cover1999elements}. As an example, $X$ could correspond to the input of a system which is transformed into a corresponding output $Y$ through a set of mathematical operations. In this case mutual information provides a quantitative measure of how efficiently the system propagates information from input $X$ to output $Y$. In most biological systems, both $X$ and $Y$ as well as the system that relates the two are inherently dynamic. Prime examples can be found, for instance, in gene regulation, where different time-varying transcription factor profiles are converted into distinct gene expression patterns through specific promoter activation- and transcription dynamics \cite{purvis2012p53, hansen2013promoter}. In these situations, $X$ and $Y$ denote complete trajectories of the input and output processes on a considered time interval $[0, t]$ and the corresponding mutual information quantifies the cumulative amount of information exchanged along these trajectories. 

While trajectory-variants of mutual information are well-established for Gaussian processes \cite{fano1961transmission, PhysRevLett.102.218101}, they remain very difficult to calculate for continuous-time Markov chains (CTMCs), which are used to describe biochemical processes involving low copy number molecules \cite{van1992stochastic}.  Recently, some progress has been made towards addressing this challenge. In \cite{9174360}, for instance, the authors derive exact expressions for the trajectory-level mutual information and channel capacity for a simple Markov chain model of transcription. For more complex systems, however, analytical solutions are generally not available and one has to resort to approximation techniques. We have previously proposed one such technique, which estimates the mutual information between complete trajectories of two molecular species by combining stochastic simulations with a moment-closure approximation of a stochastic filtering problem \cite{duso2019path}. While this approach is computationally efficient, it is so far limited to two-component systems with no additional intermediate molecular species. A related approach has been proposed in \cite{Sinzger2022}, where the authors use a generic Hawkes process approximation to solve the underlying filtering problem. 

An orthogonal method, referred to as \textit{path weight sampling} (PWS), has been developed \cite{reinhardt2022path}. The key advantages of PWS are that the resulting estimates are exact up to sampling variance and that it applies also to networks involving intermediate components. On its downside, however, it is challenging to apply PWS when the information flow between the two components $X$ and $Y$ is bidirectional (e.g., due to feedback between output and input). 

The goal of the present work is to develop an efficient and general approach to quantify mutual information between complete trajectories of any two molecular components of a chemical reaction network. To this end, we build on our previously proposed theoretical approach, but importantly, lift the assumption that no intermediate components are present. We present two Monte Carlo schemes to estimate mutual information, an exact but computationally demanding one and a much more efficient one that employs moment-closure approximations to solve the underlying filtering problem. Additionally, we propose an analytical approximation, which provides direct insight into how information transfer depends on the underlying system parameters. We use our methodology to study information processing in two archetypical network motifs, a feedforward three-component system and a positive feedback system that switches between two metastable states. Our analyses demonstrate the utility of our approach and reveal novel insights into how information propagates through networks of chemical reactions. \newline

\paragraph*{Stochastic Reaction Networks.}

We consider a well-mixed reaction network $R_Z$ consisting of $M$ chemical species $\mathrm{Z}_1,\ldots, \mathrm{Z}_M$ and $K$ reaction channels. Each reaction $k$ is defined by a stoichiometric equation
\begin{equation}
\sum_{l=1}^M \alpha_{k,l} \mathrm{Z}_l\rightarrow \sum_{l=1}^M \beta_{k,l} \mathrm{Z}_l
\end{equation}
with $\alpha_{k,l}$ and $\beta_{k,l}$ as reactant- and product multiplicities. We denote the stochastic state vector of the system by $(Z(t))_{t\geq 0}$, which tracks the copy numbers of all species over time. Each reaction channel is associated with a rate function $\lambda_k(Z(t))$, which defines how likely a reaction fires within a small amount of time given the current state of the system. Typically, $\lambda_k(Z(t))$ is given by the law of mass-action but also non-elementary rate laws such as Michaelis-Menten kinetics could be considered. When a reaction of type $k$ happens at time $t^*$, the system state changes instantaneously from $Z(t^*)$ to $Z(t^*) + \nu_k$ where $\nu_{k} = (\beta_{k,l} - \alpha_{k,l})_{l=1,\ldots, M}$. The dynamics of $Z(t)$ satisfies a Markov jump process, which can be described at the level of individual trajectories using the random time-change representation \cite{kurtz}
\begin{equation}
Z(t)=Z(0)+\sum_{k=1}^K N_k \left( \int_0^t \lambda_k(Z(s))\mathrm{d}s\right)\nu_k
\label{eq:RTC}
\end{equation}
with $N_1(t),\ldots,N_K(t)$ as independent unit Poisson processes and $Z(0)$ as the initial state of the system. We denote by $Z_{0}^t$ a complete trajectory of $Z(t)$, collecting all reaction times and types within the time interval $[0, t]$.

\paragraph*{Path Mutual Information.}
\begin{figure*}
\includegraphics[width=0.8\textwidth]{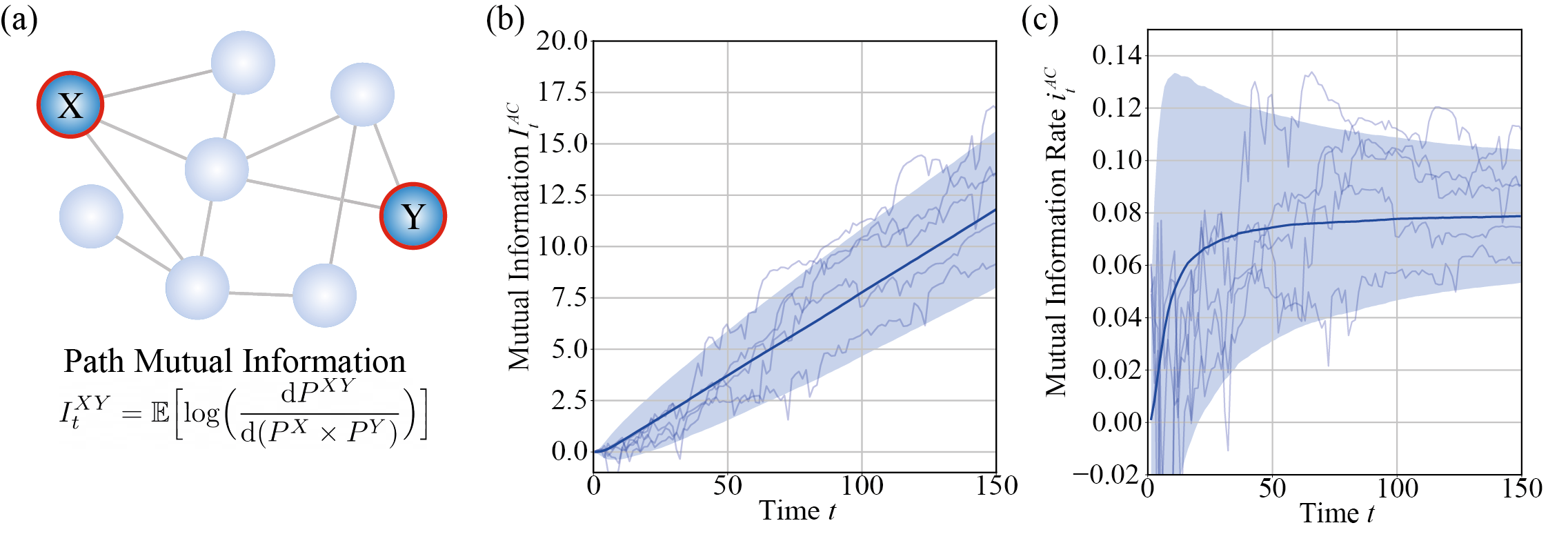}
\caption{(a) Schematic representation of a reaction network through which two molecular components $\mathrm{X}$ and $\mathrm{Y}$ exchange information. The path mutual information quantifies the cumulative amount of information exchanged between $\mathrm{X}$ and $\mathrm{Y}$ on a time interval $[0, t]$. 
(b, c) Example trajectories of the path mutual information $I_t^{XY}$ (b) and the corresponding path mutual information rate $i_t^{XY}$ (c). Calculations wer performed using reaction network (\ref{eq:ThreeNodeCRN}) with parameters $\{k_1,k_2,k_3,k_4,k_5,k_6\}=\{1,0.1,1,0.1,1,0.1\}\: \mathrm{s}^{-1}$. Mutual information rates were estimated as $i_t^{XY}\approx I_t^{XY}/t$. Thick solid lines denote averages calculated over $n=10000$ samples and shaded areas mark one standard deviation above and below the mean.}
\label{fig:fig1}
\end{figure*}
We are interested in the dynamic exchange of information between two arbitrary chemical species $\mathrm{X}=\mathrm{Z}_l$ and $\mathrm{Y}=\mathrm{Z}_j$ that are part of system (\ref{eq:RTC}) (Fig.~\ref{fig:fig1}a). To this end, we define trajectories $X_0^t \subset Z_0^t$ and $Y_0^t\subset Z_0^t$ which contain only the reaction times and types that modify $\mathrm{X}$ and $\mathrm{Y}$, respectively. For simplicity, we consider the case where $\mathrm{X}$ and $\mathrm{Y}$ do not change simultaneously. The cumulative amount of information transfer on the interval $[0, t]$ can then be quantified by the path mutual information
\begin{equation}
\label{eq:log}
I^{XY}_t=\mathbb{E}\left[ \log \frac{\mathrm{dP^{X Y}}}{\mathrm{d}(\mathrm{P}^{X}\times \mathrm{P}^{Y})}\right]
\end{equation}
with $\mathrm{P}^{X Y}$ as the \textit{joint} path measure associated with the combined trajectory $\left\{X_0^t, Y_0^t \right\}$ and $\mathrm{P}^{X}$ and $\mathrm{P}^{Y}$ as \textit{marginal} path measures corresponding to $X_0^t$ and $Y_0^t$, respectively. Note that also $\mathrm{P}^{X Y}$ is technically a marginal measure because all chemical species apart from $\mathrm{X}$ and $\mathrm{Y}$ have been integrated out. The term inside the logarithm of (\ref{eq:log}) denotes the Radon-Nikodym derivative\cite{lipster} between $\mathrm{P}^{X Y}$ and $\mathrm{P}^{X}\times\mathrm{P}^{Y}$. Evaluating the latter for paths satisfying (\ref{eq:RTC})\cite{lipster}, taking the logarithm and simplifying (Appendix Section S.1) leads to
\begin{equation}
\begin{split}
 I^{XY}_t& = \sum_{k \in R_{X} } \int_0^t \mathbb{E}\Bigl[ \lambda_k^{X Y}(s)\log (\lambda_k^{X Y}(s))  -\lambda_k^{X}(s) \log (\lambda_k^{X}(s))\Bigr]\mathrm{d}s \\
 & +  \sum_{k \in R_{Y}} \int_0^t \mathbb{E}\Bigl[\lambda_k^{X Y}(s) \log (\lambda_k^{X Y}(s))  - \lambda_k^{Y}(s)\log (\lambda_k^{Y}(s))\Bigr]\mathrm{d}s.
 %&= H_t^{Y\rightarrow X} + H_t^{X\rightarrow Y}.
 \label{eq:pathMI}
\end{split}
\end{equation}
In (\ref{eq:pathMI}), the sets $R_{X}$ and $R_{Y}$ comprise all reactions that modify $\mathrm{X}$ (or $\mathrm{Y}$), except those whose propensity depends exclusively on $X(t)$ (or $Y(t)$). Only these reactions lead ultimately to an exchange of information among $\mathrm{X}$ and $\mathrm{Y}$, which will be illustrated more concretely later in our case studies. The functions $\lambda_k^{X Y}(t)$, $\lambda_k^{X}(t)$ and $\lambda_k^{Y}(t)$ denote \textit{marginal propensities} \cite{duso2018selected, zechner2014uncoupled}, that is, the rate functions with which $(X(t), Y(t))$, $X(t)$ and $Y(t)$ evolve if the states of all other species are unknown. A marginal propensity is defined as a conditional expectation $\lambda^{A}_k(t)=\mathbb{E}[\lambda_k(Z(t))\mid A_0^t]$, corresponding to the optimal causal estimate of $\lambda_k(Z(t))$ given some partial path $A_0^t\subset Z_0^t$. This demonstrates that the amount of information transferred between sender and receiver depends on how well the sender's signal can be reconstructed from measurements taken by the receiver (and \textit{vice versa} in the presence of feedback)\cite{guo}\cite{6034225}. We remark that while only the reactions in $R_{X}$ and $R_{Y}$ show up explicitly in (\ref{eq:pathMI}), also the other reactions contribute implicitly to $I^{XY}_t$ through the inner- and outer expectations in (\ref{eq:pathMI}).

Note that the first and second line on the right hand side of Eq. (\ref{eq:pathMI}) can be identified as transfer entropies $H_t^{Y\rightarrow X}$ and $H_t^{X\rightarrow Y}$ which correspond to the fraction of information that is transferred from $\mathrm{X}$ to $\mathrm{Y}$ and from $\mathrm{Y}$ to $\mathrm{X}$, respectively. In contrast to the mutual information which by definition is symmetric in $\mathrm{X}$ and $\mathrm{Y}$, the transfer entropy provides useful insights into the causal flow of information in dynamical systems \cite{PhysRevE.95.032319}. 

In many situations, it is helpful to study information transfer at stationary states. Since $I^{XY}_t$, $H_t^{Y\rightarrow X}$ and $H_t^{X\rightarrow Y}$ generally increase with time (e.g. Fig.~\ref{fig:fig1}b), they typically diverge as $t\rightarrow \infty$. In these cases, one can resort to the corresponding \textit{information rates}, which for the quantities of interest can be defined as $i^{XY}=\lim_{t\rightarrow \infty} I^{XY}_t / t$, $h^{X\rightarrow Y}=\lim_{t\rightarrow \infty} H_t^{X\rightarrow Y}/ t$ and $h^{Y\rightarrow X}=\lim_{t\rightarrow \infty} H_t^{Y\rightarrow X} / t$ (e.g. Fig.~\ref{fig:fig1}c). 

\paragraph*{Stochastic filtering.} The central step in evaluating (\ref{eq:pathMI}) is the calculation of the conditional expectations that are required for determining the marginal propensities. In case of $\lambda_k^{X}(t)$, for instance, we have to average $\lambda_k(Z(t))=\lambda_k(\bar{z},X(t))$ with respect to the conditional probability distribution $\pi^{X}(\bar{z}, t)=P(\bar{Z}(t) = \bar{z} \mid X_0^t)$, where $\bar{Z}(t)$ is a vector containing all copy numbers of $Z(t)$ except $X(t)$. It can be shown that such a conditional probability distribution satisfies a stochastic differential equation termed a \textit{filtering equation} \cite{crisan}. In the case of $\pi^{X}(\bar{z}, t)$, this equation reads
\begin{equation}
\begin{split}
	&\mathrm{d}\pi^{X}(\bar{z}, t) \\
	&\quad= \left[\mathcal{A}^{\bar{Z}\mid X} \pi^{X}(\bar{z}, t) - \sum_{k\in R_X}(\lambda_k(\bar{z},X(t)) - \lambda_k^{X}(t)) \pi^{X}(\bar{z}, t)\right]  \mathrm{d}t\\
	& \quad + \sum_{k\in R_X}\frac{\lambda_k(\bar{z},X(t)) - \lambda_k^{X}(t)}{\lambda_k^{X}(t)}  \pi^{X}(\bar{z}, t) \mathrm{d}N_k(t),
	\label{eq:KSEQ}
\end{split}
\end{equation}
where $\mathcal{A}^{\bar{Z}\mid X}$ is an operator that is related to the generator of the original, unconditional process (see Appendix Eq. (12) for more details). Thus, in order to calculate the marginal propensity functions, we need to solve Eq. (\ref{eq:KSEQ}) and calculate the mean of $\lambda_k(\bar{z},X(t))$ using the resulting solution. 

There are two major challenges in evaluating Eq. (\ref{eq:pathMI}) in practice. First, analytical solutions of Eq. (\ref{eq:KSEQ}) and the corresponding conditional expectations are available only in exceptional cases. Second,  the outer expectation in (\ref{eq:pathMI}) is taken with respect to a distribution that is generally not known analytically. Even if that would be the case, expectations over the $x \log x$ terms in (\ref{eq:pathMI}) are most likely intractable in practice. To address these problems, we propose three different approaches which differ in scope and computational efficiency.

\paragraph*{Quasi-exact method.} This approach numerically integrates Eq. (\ref{eq:KSEQ}) on a finite-dimensional grid. This is analogous to the finite-state projection algorithm that is commonly applied to numerically solve conventional (unconditional) master equations \cite{munsky}. The outer expectation of (\ref{eq:pathMI}) is calculated as a Monte-Carlo average over $n$ independent path realizations generated using Gillespie's stochastic simulation algorithm \cite{GILLESPIE1976403}. The main advantage of this approach is that its error is fully controllable and negligible when the grid- and sample size are sufficiently large. As with other finite-state projection approaches, however, the efficiency of this approach suffers from the combinatorial explosion of states in larger reaction networks. We will use this technique to calculate ground-truth solutions for comparison with our approximate techniques described below.
\paragraph*{Moment-approximation method.} In principle, we can derive an equation for the marginal propensity $\lambda_k^X(t)$ by multiplying Eq. (\ref{eq:KSEQ}) with $\lambda_k(\bar{z},X(t))$ and summing over all $\bar{z}$. If all propensity functions are polynomial (as is the case for mass-action kinetics), this leads to a system of moment differential equations, which in general, however, is infinite-dimensional. This problem can be addressed by imposing distributional assumptions on the conditional distribution which can then be used to express moments higher than a certain order as functions of lower-order moments. While moment-closure approximations are generally ad-hoc, we found that the conditional distribution (\ref{eq:KSEQ}) is typically very well approximated by those techniques. Intuitively, this may be the case because conditional distributions are generally more informative than unconditional distributions. Throughout our case studies, we found the multivariate Gamma closure proposed in \cite{gammaclosure} to yield excellent results (Appendix Fig. 1). We remark that while the moment-approximation method requires polynomial rate functions, it may be applied also to more complex (e.g., rational) rate laws if suitable polynomial approximations are available.
\paragraph*{Analytical approximation.} Once we have obtained a closed system of conditional moment equations, we can approximate the outer expectation in (\ref{eq:pathMI}) by employing a second order Taylor expansion. In particular, if we perform an expansion around the respective expectations $\mathbb{E}[\lambda^{XY}_k(t)]=\mathbb{E}[\lambda^{X}_k(t)]=\mathbb{E}[\lambda^{Y}_k(t)]=\mathbb{E}[\lambda_k(Z(t))]$, we obtain
\begin{equation}
\begin{split}
  I^{XY}_t\approx &\sum_{k \in R_{X} } \int_0^t \frac{\mathrm{Var}[\lambda^{X Y}_k(t)]-\mathrm{Var}[\lambda^{X}_k(t)]}{2 \mathbb{E}[\lambda_k(Z(t))]} \mathrm{d}s \\ + & \sum_{k \in R_{Y}} \int_0^t \frac{\mathrm{Var}[\lambda^{X Y}_k(t)]-\mathrm{Var}[\lambda^{Y}_k(t)]}{2 \mathbb{E}[\lambda_k(Z(t))]} \mathrm{d}s.
\end{split}
\label{eq:Analytical}
\end{equation}
This equation involves variances of the marginal propensities for which approximate differential equations can be derived (see Appendix Section S.2.2). Solving these equations provides a direct way to calculate the path mutual information and its rate. 

\paragraph*{Case study (I) -- Three-Node Feedforward Network.}
The first system we want to study is a simple feedforward reaction network {\small
\begin{align} 
&\emptyset \xrightharpoonup{k_1} \mathrm{A}  				&\mathrm{A}  \xrightharpoonup{k_2 A(t)}\emptyset \nonumber\\
&\mathrm{A}  \xrightharpoonup{k_3 A(t)} \mathrm{A}+\mathrm{B}  	&\mathrm{B}  \xrightharpoonup{k_4 B(t)} \emptyset \label{eq:ThreeNodeCRN}\\
&\mathrm{B} \xrightharpoonup{k_5 B(t)} \mathrm{B}+\mathrm{C}  	&\mathrm{C}  \xrightharpoonup{k_6 C(t)}\emptyset \nonumber
\end{align}}
with $k_1,\ldots k_6$ as rate constants. We have chosen this network because it resembles an elementary motif where an input (species A) transmits information to an output (species C) through an intermediate component (species B). Moreover, all first and second-order moments are analytically tractable which will be useful to compare our results to predictions obtained from Gaussian process theory \cite{PhysRevLett.102.218101}.
Using Eq. (\ref{eq:pathMI}), the mutual information between paths $A_0^t$ and $C_0^t$ is given by
\begin{equation}
\begin{split}
I^{AC}_t &= \int_0^t  \mathbb{E}\left[k_5 \mathbb{E}\left[B(s) \mid A_0^t, C_0^t\right] \log \left(k_5 \mathbb{E}\left[B(s) \mid A_0^t, C_0^t\right]\right) \right] \mathrm{d}s \\
&\quad - \int_0^t \mathbb{E}\left[k_5 \mathbb{E}\left[B(s) \mid C_0^t \right] \log \left(k_5 \mathbb{E}\left[B(s) \mid  C_0^t\right]\right)\right] \mathrm{d}s.
\label{eq:pathMIExampleA}
\end{split}
\end{equation}
Note that (\ref{eq:pathMIExampleA}) involves terms associated with reaction $\mathrm{B}\rightarrow \mathrm{B} + \mathrm{C}$ only. This is because only through this reaction does component $C$ ultimately receive information about component $\mathrm{B}$ and hence, $\mathrm{A}$. As mentioned previously, however,  other reactions contribute implicitly to $I^{AC}_t$ through the expectation values in (\ref{eq:pathMIExampleA}).

\begin{figure*}[t]
\includegraphics[width=1\textwidth]{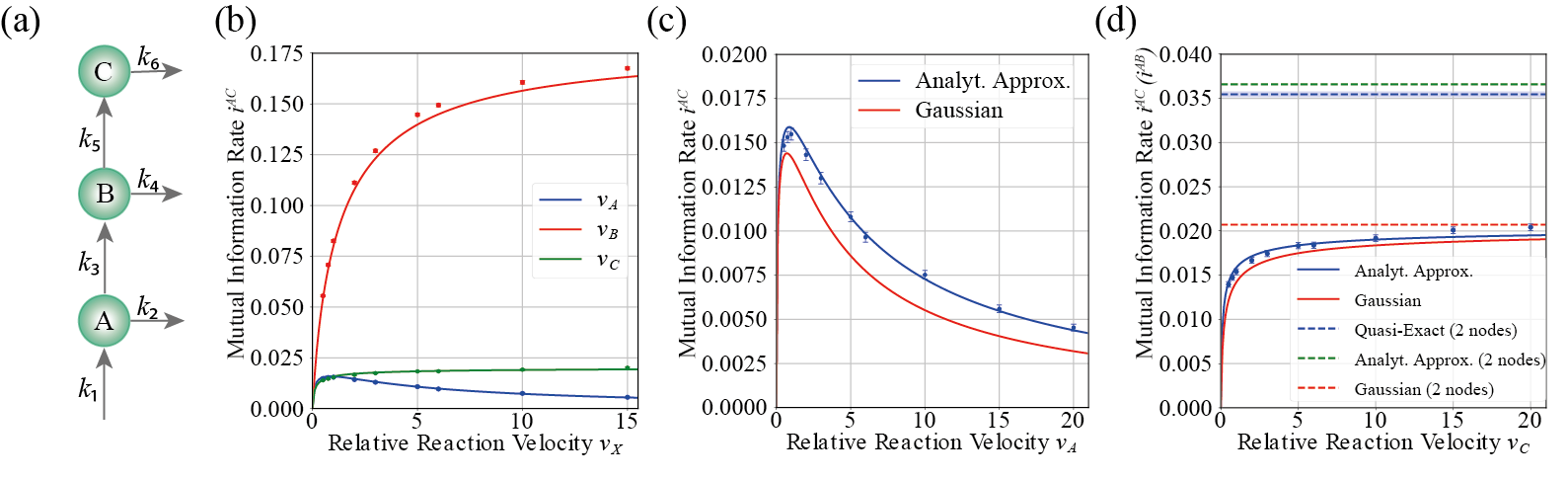}
\caption{Information transfer in a three-node feedforward network. (a) Schematic illustration of the considered network. (b) Stationary path mutual information rate between species $\mathrm{A}$ and $\mathrm{C}$ ($i^{AC}$) as a function of the relative reaction velocity $v_{\mathrm{X}}$, $X\in\{A,B,C\}$ calculated via the analytical approximation (solid lines) and the moment-approximation method (dots). (c) Stationary path mutual information rate $i^{AC}$ as a function of the relative reaction velocity $v_{\mathrm{A}}$ as in (b) and comparison to Gaussian process theory \cite{PhysRevLett.102.218101} (red). (d) Stationary path mutual information rate $i^{AC}$ as a function of the relative reaction velocity $v_{\mathrm{C}}$ as in (b) (solid lines) and comparison to the mutual information rate between nodes $\mathrm{A}$ and $\mathrm{B}$ ($i^{AB}$, dashed lines). Simulations were performed with parameters $\{k_1,k_2,k_3,k_4,k_5,k_6\}=\{1,0.1,0.1,0.1,1,0.1\}\: \mathrm{s}^{-1}$. Monte Carlo averages were calculated using $n=10000$ samples. Error bars correspond to 2.5 times the standard error.}
\label{fig:Fig2}
\end{figure*}

To study information transmission in the considered network, we calculate the stationary path mutual information rate $i^{AC}$ for different parameter regimes. To this end, we introduce reaction velocities $v_A$, $v_B$ and $v_C$ that set the time scales of production and degradation of $\mathrm{A}$, $\mathrm{B}$ and $\mathrm{C}$ without changing their average abundance. In the case of $\mathrm{A}$, for instance, we set $k_1 = \tilde{k}_1 v_A$ and $k_2 = \tilde{k}_2 v_A$ such that $\mathbb{E}[A(t)]=\tilde{k}_1 / \tilde{k}_2$ for any value of $v_A$. To verify the accuracy of our approach, we compared the moment-approximation method with the analytical- and quasi-exact methods and found good agreement among all three approaches (Fig.~\ref{fig:Fig2}b and Appendix Fig. 2).

Our analysis shows that while the mutual information rate $i^{AC}$ increases monotonically with both $v_B$ and $v_C$, it scales non-monotonically with $v_A$, exhibiting a maximum at an intermediate value of $v_A$. Intuitively, this is because for small $v_A$, species $\mathrm{B}$ is fast enough to track changes in $\mathrm{A}$ but at the same time, $\mathrm{A}$ produces little information per unit time. Correspondingly, increasing $v_A$ will initially increase $i^{AC}$ merely because more information is generated by $\mathrm{A}$. However, when $\mathrm{A}$ becomes fast in comparison to $\mathrm{B}$, a substantial amount of information is lost between $\mathrm{A}$ and $\mathrm{B}$, causing $i^{AB}$ to decrease for large $v_A$. In other words, there exists an optimal time scale of $\mathrm{A}$ that strikes a balance between the amount of information produced by $\mathrm{A}$ and the fraction of it that can be transferred to $\mathrm{C}$ via intermediate species $\mathrm{B}$. %This is in line with earlier studies that found that information transfer for a similar system is optimal  . 
In contrast, varying either $v_B$ or $v_C$ does not affect the information content in $\mathrm{A}$ but only how effectively this information can be propagated forward, leading to a simple monotonic relationship between $i^{AC}$ and $v_B$ and $v_C$, respectively.

We next use the same three-node network motif to study if and to what extent discrete-state biochemical systems differ from their continuous counterpart in terms of information transfer. To this end, we consider a real-valued variant of network (\ref{eq:ThreeNodeCRN}) where the abundances of $\mathrm{A}$, $\mathrm{B}$ and $\mathrm{C}$ are described by a chemical Langevin equation with rate functions defined in (\ref{eq:ThreeNodeCRN}). While the first and second-order moments remain identical to the discrete case, the corresponding path mutual information rate $i_{G}^{AC}$ can now be obtained from Gaussian process theory (see Appendix Section S.3.1.2). As can be proven analytically, Gaussian theory provides a lower bound on mutual information for non-Gaussian scenarios as long as the first- and second-order statistics are known \cite{mitra2001nonlinear}. This is reflected also by our analysis which compares the path mutual information rate $i^{AC}$ with $i_{G}^{AC}$ across different $v_A$ (Fig.~\ref{fig:Fig2}c).  While Gaussian theory predicts a very similar scaling and optimum of $i_{G}^{AC}$ with respect to $v_A$, it generally underestimates the information transfer between $\mathrm{A}$ and $\mathrm{C}$. 

\begin{figure*}
\includegraphics[width=0.8\textwidth]{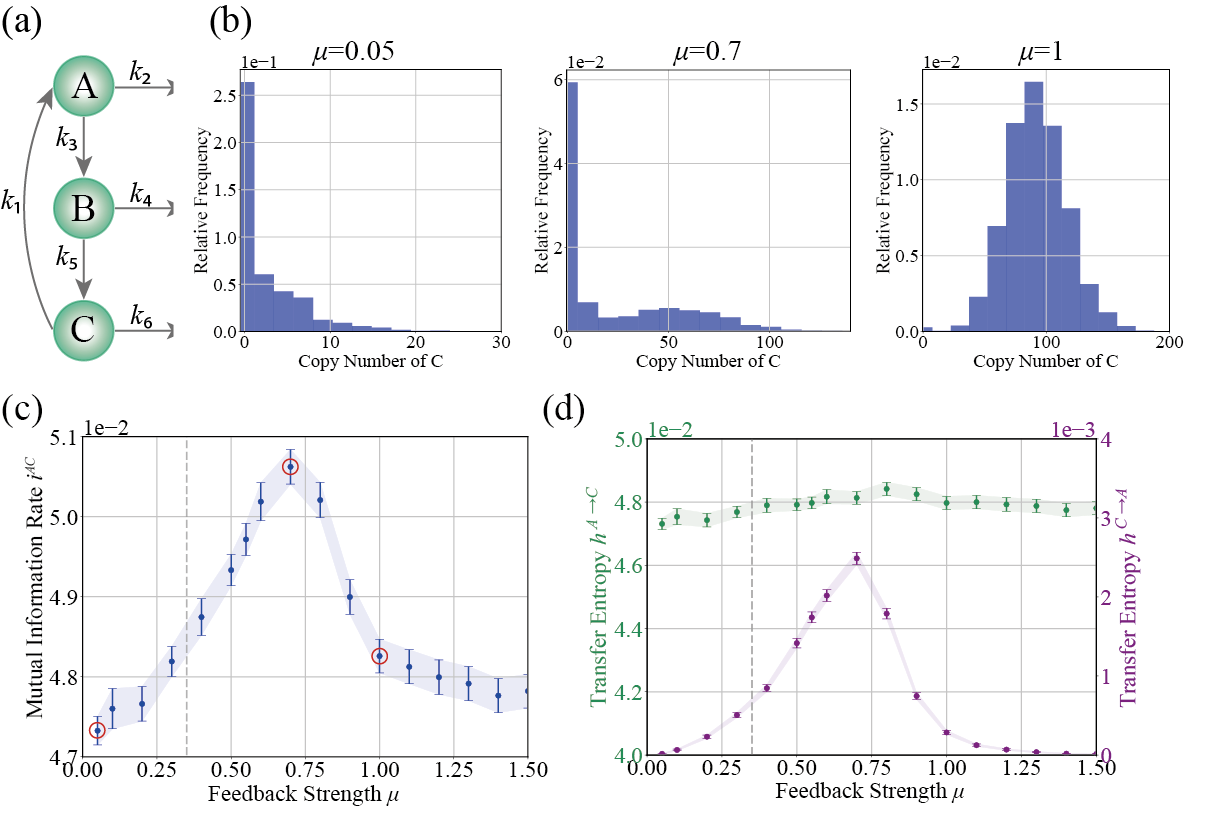}
\caption{Information transfer in a bistable system.
(a) Schematic illustration of the considered network. (b) Stationary copy number distributions of species $\mathrm{C}$ for parameters $\mu = 0.05$, $\mu = 0.7$ and $\mu = 1$.  (c) Stationary path mutual information rate $i^{AC}$ between species $\mathrm{A}$ and $\mathrm{C}$. The dashed grey line indicates the bifurcation point predicted from mean-field theory. (d) Decomposition of the path mutual information rate into forward- and backward transfer entropies $h^{A\rightarrow C}$ (green) and $h^{C\rightarrow A}$ (magenta), respectively.  Simulations were performed with the parameters $\{k_2, k_3, k_4, k_5, k_6\} = \{0.1, 1, 0.1, 0.1, 0.1\}$s$^{-1}$ and $\{K, n_H , \epsilon\}= \{30, 3, 0.03 \mathrm{s}^{-1}\}$. The system was simulated for $T=10000$ time units to reach steady state, whereas only the last $1500$ time units were used to estimate stationary information rates. Ensemble averages were calculated using $n=5000$ samples. Error bars correspond to 2.5 times the standard error.}
\label{fig:Bistable}
\end{figure*}

We next considered the limit $v_C\rightarrow \infty$ such that information in  $\mathrm{B}$  is expected to propagate to $\mathrm{C}$ in a "perfect" manner. In this case Gaussian theory predicts that 
\begin{equation}
	\lim_{v_C \rightarrow \infty} i^{AC}_G= -\frac{k_2}{2}+\frac{1}{2}\sqrt{k_2(k_2+k_3)},
	\label{eq:GaussLimit}
\end{equation}
which coincides with the Gaussian mutual information rate between $\mathrm{A}$ and $\mathrm{B}$, $i_G^{AB}$. In other words, the three-node network reduces to a two-node network for Gaussian processes when $v_C \rightarrow \infty$, as might be expected intuitively. To compare these results with the discrete-state network, we determined $\lim_{v_C \rightarrow \infty} i^{AC}$ analytically based on (\ref{eq:Analytical}), which happens to coincide exactly with (\ref{eq:GaussLimit}) (see Appendix Section S.3.1 for a derivation). In case of the discrete-state system, however,  this asymptotic limit is lower than the mutual information rate between $\mathrm{A}$ and $\mathrm{B}$, which based on (\ref{eq:Analytical}) is approximated as 
\begin{equation}
\label{eq:lim1}
i^{AB}\approx-\frac{k_2}{2}+\frac{1}{2}\sqrt{k_2(k_2+2k_3)}.
\end{equation}
Numerical simulations show that both $i^{AB}$ and $i^{AC}$ are approximated accurately through (\ref{eq:Analytical}) (Fig.~\ref{fig:Fig2}d and Appendix Fig. 2c). Calculating the ratio between $i^{AB}$ and $i^{AC}$ and taking the respective limits with respect to $k_2$ suggests that $i^{AC}$ is lower than $i^{AB}$ by a factor of at least $\sqrt{2}$ ($k_2 \rightarrow 0$) and at most $2$ ($k_2 \rightarrow \infty$) for any $k_3 > 0$. This demonstrates that in the discrete-state scenario, a certain amount of information is inevitably lost when two species $\mathrm{A}$ and $\mathrm{C}$ communicate through an intermediate species $\mathrm{B}$, even when $v_C\rightarrow \infty$. This is in stark contrast to the continuous-state scenario where the amount of information lost through intermediate species $\mathrm{B}$ can be made arbitrarily small by increasing $v_C$. In summary, our results show that the discrete nature of biochemical systems can lead to not only quantitative, but even qualitative differences when compared to continuous-state systems. \newline
\paragraph*{Case study (II) -- Bistable switch.}
As a second example, we consider a variant of the previous network that exhibits more complex non-linear dynamics. In particular, we introduce positive feedback between species $\mathrm{C}$ and $\mathrm{A}$ by replacing the constant production rate of $\mathrm{A}$ with one that increases with the abundance of $\mathrm{C}$ (Fig.~\ref{fig:Bistable}a). More precisely, we choose a Hill-type rate law of the form
\begin{equation}
 k_1(C(t))=\mu \frac{C(t)^{n_H}}{K^{n_H}+C(t)^{n_H}}+\epsilon
\end{equation}
with $\mu$, $n_H$, $K$ and $\epsilon$ as positive constants. For certain parameter regimes, this system is bistable where individual trajectories switch stochastically between two modes (see caption of Fig.~\ref{fig:Bistable} for specific parameter values). 
The goal of this case study is to understand how such bistability affects information transfer in biochemical systems. To this end, we applied our moment-approximation method to estimate stationary mutual information rates between species $\mathrm{A}$ and $\mathrm{C}$.  

We first analyzed how the mutual information rate changes for varying feedback strengths $\mu$. For low and high values of $\mu$, the system exhibits a single mode, while intermediate values of $\mu$ lead to bimodal behavior (Fig.~\ref{fig:Bistable}b). This is resembled qualitatively by a corresponding mean-field model (Appendix Fig.~3). Fig.~\ref{fig:Bistable}c shows that before the bifurcation point, the mutual information rate remains nearly constant with increasing $\mu$ but then begins to increase until it reaches a certain maximum. At this point, the system fluctuates between two equilibrium points as can be seen from the copy numbers of the species being bimodally distributed (Fig.~\ref{fig:Bistable}b). Beyond this maximum, the bimodality becomes less pronounced and $i^{AC}$ decreases again. Interestingly, this suggests that information transfer is most effective in regimes where the system as a whole is very noisy. 

To understand this better, we decomposed the mutual information rate into the transfer entropy rates $h^{A\rightarrow C}$ and $h^{C\rightarrow A}$, quantifying the causal flow of information from $\mathrm{A}$ to $\mathrm{C}$ and from $\mathrm{C}$ to $\mathrm{A}$, respectively (Fig.~\ref{fig:Bistable}d). Interestingly, this shows that the forward contribution $h^{A\rightarrow C}$ is more or less the same for all considered feedback strengths $\mu$, regardless of whether the system exhibits one or two modes. 
By contrast, the backward contribution $h^{C\rightarrow A}$ is approximately zero for very small $\mu$, but then shows a peak that is located within the bimodal regime. From this point on, the second equilibrium becomes more and more populated and the backward contribution $h^{C\rightarrow A}$ decreases again. For very large $\mu$, the rate $k_1$ is strongly saturated such that the system effectively reduces to the simpler feedforward motif discussed in the previous section, where no information flows along the backward direction $\mathrm{A}\rightarrow \mathrm{C}$. 

In summary, this shows that information transfer between two molecular species can be significantly enhanced by positive feedback. This is the case when the feedback strength is in a regime where it generates multiple meta-stable equilibria that the system can attain. In this situation, both the forward- and backward contribution to the mutual information are significantly different from zero, leading to large values overall. More generally, this analysis illustrates the applicability of our approach to complex and strongly nonlinear dynamical systems that are beyond Gaussian theory.
\paragraph*{Conclusions.}
In this work we have developed a general method to quantify information transmission in biochemical networks via the path mutual information. This method exploits a fundamental relationship between mutual information and filtering theory. We have first introduced a quasi-exact Monte Carlo scheme that combines conventional stochastic simulations with a brute-force numerical solution of the underlying filtering equations. While this approach was needed to calculate ground-truth solutions, it quickly becomes computationally infeasible as the system size grows. As we have shown in our earlier work \cite{duso2019path}, this problem can be addressed using moment-closure approximations that project the filtering distribution onto a finite set of (approximate) moments, which are obtained by solving a system of differential equations. In our numerical experiments, we found the approximate Monte Carlo method based on the Gamma-closure to be in very good agreement with the exact path mutual information, although different closures may be required for other types of systems.

We have further shown how the outer expectation in the calculation of the path mutual information can be approximated analytically. In this way, Monte Carlo sampling can be avoided entirely and the path mutual information becomes analytically accessible. Although the proposed approximation appears relatively coarse, it was in surprisingly close agreement with Monte Carlo estimation. Deriving similar approximations in a more principled manner will be an interesting avenue for future research.

When applied to our case studies, the path mutual information revealed interesting insights into how information propagates across cascades of chemical reactions. For instance, we found that the discreteness of chemical processes leads to quantitative but even qualitative differences in information transmission when compared to equivalent processes defined on a continuous state space -- even when their first- and second order statistics are identical. In our second case study, we have studied information transfer in a non-linear, positive feedback system. Our analysis revealed that positive feedback -- and the resulting bistability -- can enhance information transmission between input and output. By decomposing the mutual information into the respective transfer entropies, we found that this enhancement is due to an increased backward-contribution to the mutual information (i.e., from output to input) while the forward-contribution remains largely unaffected by the presence of feedback. Interestingly, the backward contribution is maximal in the bimodal regime, when the system switches randomly and evenly between the two modes.

In summary, our results highlight the need for information theoretical concepts that are compatible with the discrete- and nonlinear dynamics of biochemical networks. The methodology outlined in this work aims to fill this gap and we envision several interesting applications in the future. For instance, it could be used to identify network architectures and parameter regimes that are optimal in terms information transfer and understand how those compare to evolved intracellular systems. Beyond this, mutual information- and transfer entropy rates play an important role in the context of stochastic thermodynamics, for instance to derive second-law-like inequalities for feedback-controlled systems \cite{horowitz2014second}. Understanding how the information processing capabilities of biochemical systems are limited by thermodynamic constraints will be an interesting subject for future work.

\subsection*{CODE AVAILABILITY}
Python code underlying all our simulations is available at https://github.com/zechnerlab/PathMutualInformation/.

\begin{acknowledgments}
The authors thank Carl Modes and Tommaso Bianucci for useful insights and critical feedback on the work. This work was supported by core funding of the Max Planck Institute of Molecular Cell Biology and Genetics.
\end{acknowledgments}

%\nocite{*}
%\bibliography{aipsamp}% Produces the bibliography via BibTeX.
%apsrev4-2.bst 2019-01-14 (MD) hand-edited version of apsrev4-1.bst
%Control: key (0)
%Control: author (8) initials jnrlst
%Control: editor formatted (1) identically to author
%Control: production of article title (0) allowed
%Control: page (0) single
%Control: year (1) truncated
%Control: production of eprint (0) enabled
%

\end{document}

\maketitle
\section{Derivation of the Path Mutual Information}
The following derivation is based on our previous work \cite{duso2019path}. As discussed in the main text, the mutual information between two trajectories $X_0^t \subset Z_0^t$ and $Y_0^t \subset Z_0^t$ can be defined as
\begin{equation}
\label{eq:MI}
I_t^{XY}=\mathbb{E}\left[ \log \frac{\mathrm{dP^{X Y}}}{\mathrm{d}(\mathrm{P}^{X}\times \mathrm{P}^{Y})}\right] %= \int \log \frac{\mathrm{dP^{X Y}}}{\mathrm{d}(\mathrm{P}^{X}\times \mathrm{P}^{Y})} \mathrm{d}\mathrm{P}^{XY}
\end{equation}
where $\mathrm{dP^{X Y}}/\mathrm{d}(\mathrm{P}^{X}\times \mathrm{P}^{Y})$ is the Radon-Nikodym derivative between the joint path measure $\mathrm{P}^{X Y}$ and the product of the \textit{marginal} path measures $\mathrm{P}^{X}$ and $\mathrm{P}^{Y}$. Intuitively, the Radon-Nikodym derivative corresponds to a likelihood ratio between two competing probability laws, in this case $\mathrm{P}^{X Y}$ and $\mathrm{P}^{X} \times\mathrm{P}^{Y}$, respectively.  In other words, it assesses how much more likely a joint path $\{X_0^t, Y_0^t\}$ originates from the (true) joint probability measure $\mathrm{P}^{X Y}$ than from the product measure $\mathrm{P}^{X} \times \mathrm{P}^{Y}$ in which $X_0^t$ and $Y_0^t$ are considered independent. 

We remark that all three probability measures $\mathrm{P}^{X Y}$, $\mathrm{P}^{X}$ and $\mathrm{P}^{Y}$ are marginal measures because all remaining components are ''integrated out''.  In other words, they are the probability laws that capture how $(X(t), Y(t))$, $X(t)$ and $Y(t)$ evolve if no knowledge about all other components in $Z(t)$ is available. We and others have previously shown how the dynamics of such marginal processes can be obtained using the theory of stochastic filtering \cite{zechner2014uncoupled}\cite{duso2018selected}. To illustrate this, we focus on the case of $Y(t)$ but $(X(t), Y(t))$ and $X(t)$ can be handled analogously. If knowledge about the whole state $Z(t)$ is available, the dynamics of $Y(t)$ satisfies
\begin{equation}
	Y(t) = Y(0) + \sum_{k\in R_{\bar{Y}}} U_k\left( \int_0^t \lambda_k(Z(t)) \right) v_k^Y,
	\label{eq:condY}
\end{equation}
where $v_k^Y$ is the element of the stoichiometric change vector $v_k$ that acts upon $Y(t)$.  With $R_{\bar{Y}}$ and $R_{\bar{X}}$ we denote the set of reactions that modify $Y(t)$ and $X(t)$, respectively. Note that we focus here on the case where there are no reactions that change $X(t)$ and $Y(t)$ simultaneously (e.g., $\mathrm{X}\rightarrow \mathrm{Y}$) such that $R_{\bar{X}}$ and $R_{\bar{Y}}$ are disjoint sets. As can be seen from (\ref{eq:condY}), the dynamics of $Y(t)$ depends on the complete state $Z(t)$ and as such is not self-contained. In more technical terms, (\ref{eq:condY}) describes the dynamics of $Y(t)$ relative to the natural filtration of $Z(t)$, that is, the complete history of events $Z_0^t$ that brought the full system from $Z(0)$ to $Z(t)$. To derive the marginal dynamics of $Y(t)$, one then requires $Y(t)$ to depend no longer on the complete history $Z_0^t$, but only on its own history $Y_0^t \subset Z_0^t$. The latter contains information about all reactions $R_{\bar{Y}}$ that modify species $Y(t)$. The innovation theorem \cite{aalen2008} then states that the dynamics of $Y(t)$ relative to $Y_0^t$ satisfies
\begin{equation}
	Y(t) = Y(0) + \sum_{k\in R_{\bar{Y}}} U_k\left( \int_0^t \lambda_k^Y(t) \right) v_k^Y,
	\label{eq:margY}
\end{equation}
with $\lambda_k^Y(t) = \mathbb{E}[\lambda_k(Z(t)) \mid Y_0^t]$. In other words, the original propensities are replaced by their expectation conditionally on $Y_0^t$, which we refer to as \textit{marginal propensity}. Importantly, Eq. (\ref{eq:margY}) is now self-contained, because the dependency on all components except $Y(t)$ has been integrated out through the conditional expectation. We remark that (\ref{eq:margY}) is exact: solving (\ref{eq:margY}) -- for instance through stochastic simulation -- will generate paths $Y_0^t$ consistent with the marginal probability measure $\mathrm{P}^Y$. Analogous constructions can be performed for $\mathrm{P}^{XY}$ and $\mathrm{P}^X$, respectively.

To determine the Radon-Nikodym derivative in (\ref{eq:MI}), we focus on the joint trajectories $\{X_0^t, Y_0^t\}$, which contain information about all reactions that modify $X(t)$ and $Y(t)$. We can now employ Jacod's formula for the Radon-Nikodym derivative \cite{lipster}, which for a counting process $\Xi(t)$ consisting of $K$ counting processes (e.g., such as Eq. (2) in the main text) and with known initial state $\Xi(0)$ takes the form
\begin{equation}
\label{eq:genrnd}
\frac{\mathrm{dW}}{\mathrm{dQ}}= \frac{\prod_{k\in K}(\prod_{j=1}^{N_k(t)}\phi_k(\tau_{k,j}))\exp(-\int_0^t \phi_k(s) \mathrm{d}s)}{\prod_{k\in K}(\prod_{j=1}^{N_k(t)}\tilde{\phi}_k(\tau_{k,j}))\exp(-\int_0^t \tilde{\phi}_k(s) \mathrm{d}s)},
\end{equation}
where $\mathrm{W}$ and $\mathrm{Q}$ are the measures under which the process has propensity functions $\phi_k$ and $\tilde{\phi}_k$, respectively. Note that $\phi_k$ and $\tilde{\phi}_k$ can in general depend on the state of $\Xi(t)$ or even the whole process history $\Xi_0^t$. The symbol $\tau_{k, j}$ denotes the time point right before the $j$th reaction of type $k$ happens. Instantiating (\ref{eq:genrnd}) for the derived marginal processes (e.g., Eq. (\ref{eq:margY}) in the case of $Y(t)$) yields
\begin{equation}
\label{eq:rnd}
\frac{\mathrm{dP}^{X Y}}{\mathrm{d}(\mathrm{P}^{X}\times \mathrm{P}^{Y})}  = \frac{\prod_{k \in R_{\tilde{X}}\cup R_{\tilde{Y}}} (\prod_{j=1}^{N_k(t)} \lambda_k^{X Y}(\tau_{k,j}))\exp (-\int_0^t \lambda_k^{X Y}(s)\mathrm{d}s) }{\prod_{k \in R_{\tilde{X}}}(\prod_{j=1}^{N_k(t)} \lambda_k^{X}(\tau_{k,j}))\exp (-\int_0^t \lambda_k^{X}(s)\mathrm{d}s)\times \prod_{k \in R_{\tilde{Y}}}(\prod_{j=1}^{N_k(t)} \lambda_k^{Y}(\tau_{k,j}))\exp (-\int_0^t \lambda_k^{Y}(s)\mathrm{d}s) }
\end{equation}
with $\lambda_k^{X Y}(t) = \mathbb{E}[\lambda_k(Z(t)) \mid X_0^t, Y_0^t]$, $\lambda_k^{X }(t) = \mathbb{E}[\lambda_k(Z(t)) \mid X_0^t]$ and $\lambda_k^{Y}(t) = \mathbb{E}[\lambda_k(Z(t)) \mid Y_0^t]$ as the marginal propensities of processes $(X(t), Y(t))$, $X(t)$ and $Y(t)$. Eq. (\ref{eq:rnd}) can be further simplified by realizing that reactions in $R_{\bar{X}}$ for which $\lambda_k^{X Y}(t) = \lambda_k^{X}(t)$ cancel out (and equivalently for $R_{\bar{Y}}$). This is the case for reactions whose propensity function depends exclusively on $X(t)$ (or $Y(t)$). Therefore, we only have to consider reactions in $R_{\bar{X}}$ (or $R_{\bar{Y}}$) that depend on species other than $X(t)$ (or $Y(t)$) such as $\mathrm{Z}_k \rightarrow \mathrm{Z}_k + \mathrm{X}$. We refer to these reactions as $R_{X}$ and $R_{Y}$, respectively.  The set of reactions in which species $X$ gets affected by an arbitrary species $Z_l \neq X$ is denoted as $R_{X} \subset R_{\tilde{X}}$. As an example, a catalytic reaction $\mathrm{Z}_k \rightarrow \mathrm{Z}_k + \mathrm{X}$ would be included in $R_X$ whereas a first-order degradation reaction $\mathrm{X}\rightarrow \emptyset$ would be contained in $R_{\tilde{X}}$ but not $R_X$. With these definitions, and realizing that $\prod_{j=1}^{N_k(t)} f(\tau_{k, j})= \exp(\int_0^t \log f(s) \mathrm{d}N_k(s))$ yields for the path mutual information 
\begin{equation}
\label{eq:prpathMI}
\begin{split}
 I_t^{XY} &= \mathbb{E} \Bigl[\sum_{k \in R_X} \Bigl( \int_0^t \bigl( \log( \lambda_k^{X Y}(s))-\log( \lambda_k^{X}(s))\bigr) \mathrm{d}N_k(s) -\int_0^t \bigl( \lambda_k^{X Y}(s)-\lambda_k^{X}(s)\bigr) \mathrm{d}s\Bigr)\\ &\quad+  \sum_{k \in R_Y} \Bigl( \int_0^t \bigl( \log (\lambda_k^{X Y}(s))-\log (\lambda_k^{Y}(s))\bigr) \mathrm{d}N_k(s) - \int_0^t \bigl( \lambda_k^{X Y}(s)-\lambda_k^{Y}(s)\bigr) \mathrm{d}s\Bigr) \Bigr] .
\end{split}
\end{equation}
Moreover, we can decompose the reaction counters $N_k(t)$ into a predictable part and a martingale such that $\mathrm{d}N_k(t) = \lambda_k^{XY}(t)\mathrm{d}t + \mathrm{d}\tilde{N}_k(t)$, where $\mathrm{d}\tilde{N}_k(t)$ is a \textit{centered} process which is zero on average. In addition, we observe that 
\begin{equation}
\label{eq:intout}
\begin{split}
\mathbb{E}[\lambda_k^X (t)] & =\mathbb{E}[\mathbb{E}[\lambda_k(Z(t)) \mid X_0^t]]=%\int \sum_{\bar{z}}\lambda_k(\bar{z}, X(t))  P(\bar{Z}(t)= \bar{z}\mid X_0^t)\mathrm{d}P^X \\ & =\sum_{x,\bar{z}} \lambda^Z (t)P(\bar{Z}(t)=\bar{z},X(t)=x)= \sum_z \lambda^Z (t) P(Z(t)=z) =
\mathbb{E}[\lambda_k(Z(t))].
\end{split}
\end{equation}
Then, by changing the order of integration and expectation, we further obtain
\begin{equation}
\begin{split}
 I_t^{XY} &= \sum_{k \in R_X} \int_0^t \mathbb{E} \Bigl[ \left(  \log( \lambda_k^{X Y}(s))-\log( \lambda_k^{X}(s))\right) \left(\lambda_k^{XY}(s)\mathrm{d}s + \mathrm{d}\tilde{N}(s)\right)\Bigr] -\int_0^t \mathbb{E} \Bigl[ \lambda_k^{X Y}(s)-\lambda_k^{X}(s)\Bigr] \mathrm{d}s\\ &\quad+  \sum_{k \in R_Y} \int_0^t \mathbb{E} \Bigl[ \left( \log (\lambda_k^{X Y}(s))-\log (\lambda_k^{Y}(s))\right)  \left(\lambda_k^{XY}(s)\mathrm{d}s + \mathrm{d}\tilde{N}(s)\right)\Bigr]  - \int_0^t \mathbb{E} \Bigl[ \lambda_k^{X Y}(s)-\lambda_k^{Y}(s)\Bigr] \mathrm{d}s \\
 &=  \sum_{k \in R_X} \int_0^t \mathbb{E} \Bigl[ \left(  \log( \lambda_k^{X Y}(s))-\log( \lambda_k^{X}(s))\right) \lambda_k^{XY}(s) \Bigr] \mathrm{d}s \\ &\quad+  \sum_{k \in R_Y} \int_0^t \mathbb{E} \Bigl[ \left( \log (\lambda_k^{X Y}(s))-\log (\lambda_k^{Y}(s))\right)  \lambda_k^{XY}(s)\Bigr] \mathrm{d}s ,
\end{split}
\end{equation}
where the second equality follows from the martingality of $\tilde{N}_k(t)$ and from $\mathbb{E} [ \lambda_k^{XY}(t) ] = \mathbb{E} [ \lambda_k^{X}(t) ] = \mathbb{E} [ \lambda_k(Z(t)) ]$ (see Eq. (\ref{eq:intout})). Finally, by realizing that 
\begin{equation}
\begin{split}
\mathbb{E}[\lambda^{XY}(t) \log (\lambda^X (t))] & = \mathbb{E}[\mathbb{E}[\lambda^{XY}(t) \log (\lambda^X (t)) \mid X_0^t]] \\
&=\mathbb{E}[\underbrace{\mathbb{E}[\lambda^{XY}(t) \mid X_0^t]}_{\lambda_k^X(t)} \log (\lambda^X (t)) ] \\ 
& = \mathbb{E}[\lambda_k^{X}(t) \log (\lambda_k^X (t))],
\end{split}
\end{equation}
we arrive at
\begin{equation}
\label{eq:finalpathmi}
\begin{split}
 I_t^{XY} & = \sum_{k\in R_X} \int_0^t \mathbb{E}\Bigl[ \lambda_k^{X Y}(s)\log (\lambda_k^{X Y}(s)) -\lambda_k^{X}(s) \log (\lambda_k^{X}(s))\Bigr]\mathrm{d}s\\ & \quad+  \sum_{k \in R_Y} \int_0^t \mathbb{E}\Bigl[\lambda_k^{X Y}(s) \log (\lambda_k^{X Y}(s))- \lambda_k^{Y}(s)\log (\lambda_k^{Y}(s))\Bigr]\mathrm{d}s. 
\end{split}
\end{equation}
Note that even though Eq.~(\ref{eq:finalpathmi}) is compact and beneficial for analytical purposes, it is numerically more difficult to handle than Eq. (\ref{eq:prpathMI}). In particular, it would require evaluating the marginal propensities, calculating Monte Carlo averages over the corresponding $x \log x$ terms and numerically integrating the resulting averages over time. Using (\ref{eq:prpathMI}), by contrast, one first calculates the time-integral (for which effective numerical solvers can be used) and subsequently averages over many Monte Carlo samples. We therefore use Eq. (\ref{eq:prpathMI}) instead for all our numerical simulations.

\section{Stochastic filtering}
To calculate mutual information between paths $X_0^t$ and $Y_0^t$, we require the marginal propensities $\lambda_k^{XY}(t)$, $\lambda_k^{X}(t)$ and $\lambda_k^{Y}(t)$ as we have seen in the previous section. We will from now on focus on the calculation of $\lambda_k^{X}(t)$, but remark that $\lambda_k^{XY}(t)$ and $\lambda_k^{Y}(t)$ can be obtained analogously. Recall that a marginal propensity is defined as 
\begin{equation}
\lambda_k^{X}(t) = \mathbb{E}[\lambda_k(Z(t)) \mid X_0^t] = \mathbb{E}[\lambda_k(\bar{Z}(t), Y(t)) \mid X_0^t] = \sum_{\bar{z}} \lambda_k(\bar{z}, X(t))\pi^X(\bar{z}, t)
\end{equation}
where $\bar{Z}(t)$ is a vector collecting all molecular abundances except $X(t)$. The average is taken with respect to a conditional probability distribution $\pi^X(\bar{z}, t) := P(\bar{Z}(t) = \bar{z} \mid X_0^t)$, also referred to as \textit{filtering equation}. While a detailed and more general derivation of this equation can be found in \cite{duso2018selected}, we will here focus on the particular form of this equation required for the considered systems. In particular, it can be shown that $\pi^X(\bar{z}, t)$ satisfies the stochastic differential equation
\begin{equation}
\label{eq:filtereq}
\begin{split}
\mathrm{d}\pi^{X}(\bar{z},t) & = \underbrace{\sum_{k\in R_{\bar{Z}}}\left[\lambda_k(\bar{z}-v_k^{\bar{Z}}, X(t))\pi^{X}(\bar{z}-v_k^{\bar{Z}},t)-\lambda_k(\bar{z}, X(t))\pi^{X}(\bar{z},t)\right]}_{\mathcal{A}^{\bar{Z}\mid X}\pi^{X}(\bar{z},t) }\mathrm{d}t \\
&\quad - \sum_{k\in R_{X}}(\lambda_k(\bar{z}, X(t))-\lambda_k^{X}(t) )\pi^{X}(\bar{z},t)\mathrm{d}t  + \sum_{k\in R_{X}} \frac{\lambda_k(\bar{z}, X(t))-\lambda_k^{X}(t)}{\lambda_k^{X}(t)}\pi^{X}(\bar{z},t) \mathrm{d}N_k(t),
\end{split}
\end{equation}
where $v_k^{\bar{Z}}$ is the part of the stoichiometric change vector $v_k$ that acts on $\bar{Z}(t)$ and $\mathrm{d}N_k(t)$ is the differential version of the counting process $N_k(t)$, which is one exactly at the time points where reaction $k$ happens and zero otherwise. The reaction set $R_{\bar{Z}}$ contains all reaction indices that modify components in $\bar{Z}(t)$ while $R_{X}$ contains all reactions that modify $X(t)$ but whose rates depend on components in $\bar{Z}(t)$ (e.g., $\mathrm{Z}_i \rightarrow \mathrm{Z}_i + \mathrm{X}$). Note that for simplicity we assume that reactions in $R_{X}$ modify exclusively the copy number of $\mathrm{X}$, while species $\bar{\mathrm{Z}}$ are neither consumed nor degraded. 

The terms in the first row of (\ref{eq:filtereq}) can be summarized as $\mathcal{A}^{\bar{Z}\mid X}\pi^{X}(\bar{z},t)$, where $\mathcal{A}^{\bar{Z}\mid X}$ is a forward operator acting on $\pi^{X}(\bar{z},t)$. This part is essentially the same as what we encounter on the right hand side of a conventional master equation with state $\bar{z}$ and rates $\lambda_k(\bar{z}, X(t))$ for all $k\in R_{\bar{Z}}$ that depend not only on the internal state $\bar{z}$, but also on some external process $X(t)$. More specifically, it describes how the components $\bar{Z}(t)$ evolve as if the reactions in $R_X$ were switched off. The second row accounts for the knowledge about $\bar{Z}(t)$ that is gained through observing (or not observing) reactions in $R_X$. Concrete examples will be given later in this appendix.

\subsection{Moment approximations}

Direct numerical solution of (\ref{eq:filtereq}) is computationally demanding due to the combinatorial explosion of states in larger reaction networks. More effective solutions of (\ref{eq:filtereq}) can be obtained using moment-closure techniques as we have shown also previously \cite{duso2019path, zechner2014uncoupled, duso2018selected}. This technique is suitable when the reaction propensities are of polynomial form such as encountered with mass-action kinetics. If this is the case, Eq. (\ref{eq:filtereq}) can be readily transformed into an equivalent system of moment equations, which is subsequently truncated to obtain a finite-dimensional moment hierarchy. The first step can be achieved by multiplying (\ref{eq:filtereq}) with a polynomial in $\bar{z}$ (corresponding to a certain desired moment) and summing over all possible values of $\bar{z}$. This generally leads to an infinite-dimensional system of moment equations which needs to be truncated at a certain order. This can be achieved by assuming the underlying distribution $\pi^{X}(\bar{z},t)$ to belong to a certain family of probability distributions, which can be described by a finite (and small) number of degrees of freedom (e.g., mean $\mu$ and standard deviation $\sigma$ in case of a univariate Gaussian distribution). This can then be exploited to express higher-order moments as a function of lower-order moments, leading to a closed system of differential equations. In our previous works \cite{duso2019path, zechner2014uncoupled}, we have used a third-order Gamma-closure to approximate conditional moments of a one-dimensional filtering equation and have found very good accuracy. Since we consider multidimensional filtering equations in this work, we analogously use a multivariate extension of the third order Gamma-closure \cite{gammaclosure}, for which
\begin{equation}
\mathbb{E}[\bar{Z}_j^2 \bar{Z}_l \mid X_0^t] = 2\frac{\mathbb{E}[\bar{Z}_j^2\mid X_0^t]\mathbb{E}[\bar{Z}_j \bar{Z}_l\mid X_0^t]}{\mathbb{E}[\bar{Z}_j\mid X_0^t]}- \mathbb{E}[\bar{Z}_j^2\mid X_0^t]\mathbb{E}[\bar{Z}_l \mid X_0^t].
\label{eq:GammaClosure}
\end{equation}
Note that (\ref{eq:GammaClosure}) agrees with the univariate Gamma-closure if $j=l$.  A concrete application of the multivariate Gamma-closure will be presented later in Section 3.1.1.

\begin{figure}
\centering
\includegraphics[width=0.3\textwidth]{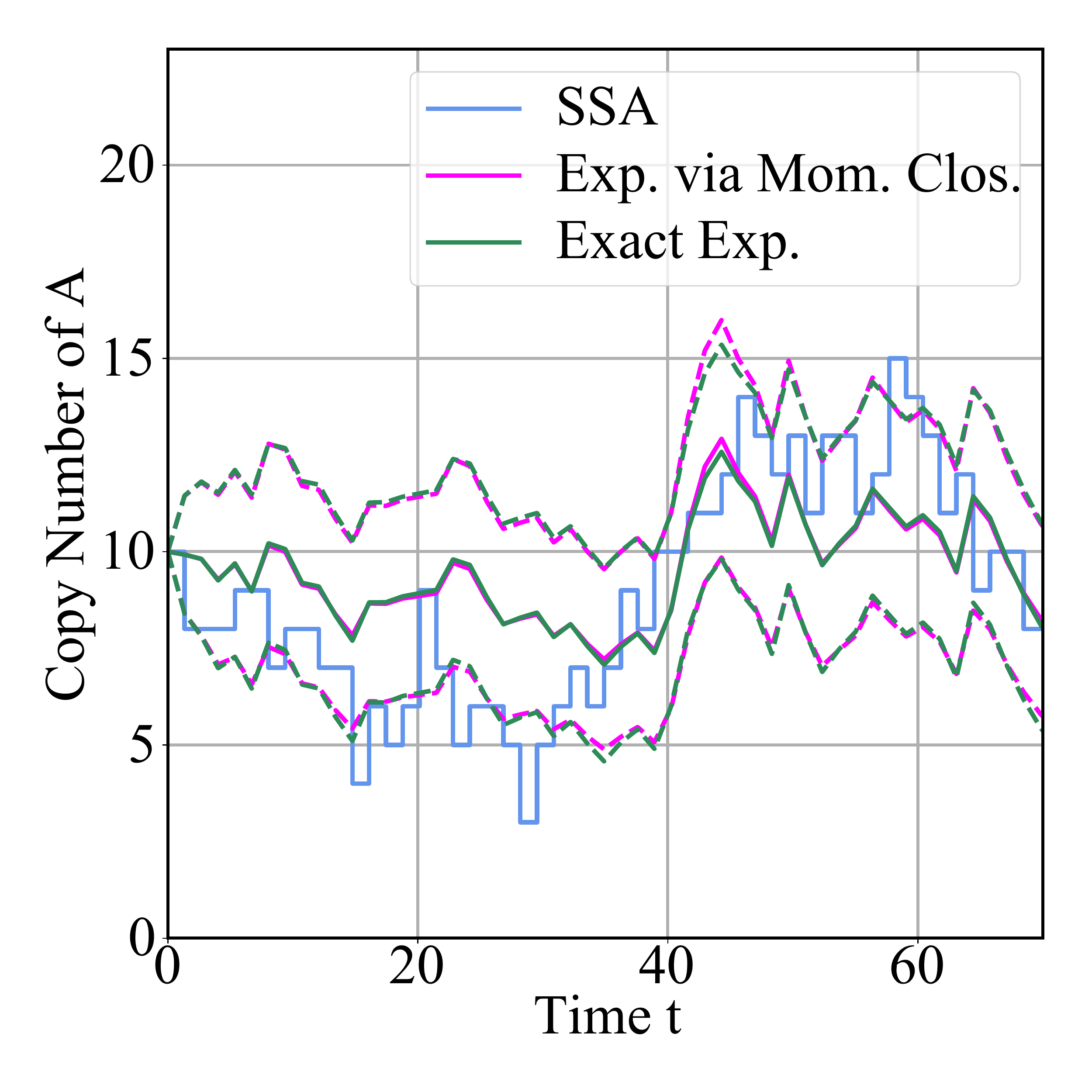}
\caption{Solution of the filtering equation obtained by direct integration on a finite grid (\textit{quasi-exact method}) and moment-closure (\textit{moment approximation method}). The considered example is a two-node feedforward network, for which we calculated $P(A(t) = a \mid B_0^t)$ (see Section 3.1.1). Shown are means (solid lines) and standard deviations (dashed lines) as well as the underlying realization of $A(t)$ (blue stair plot). Parameters were set to $\{k_1,k_2,k_3,k_4\}=\{1,0.1,0.1,0.1\}\: \mathrm{s}^{-1}$ and $\{a_0,b_0\}=\{10,10 \}$.}
\label{fig:et}
\end{figure}

Figure \ref{fig:et} shows example trajectories of the first and second conditional moments obtained by direct numerical integration of (\ref{eq:filtereq}) and the Gamma-closure. 
\subsection{Analytical Approximation}
In this section, we derive a simple analytical approximation of the path mutual information as shown in Eq. (6) in the main text. For that sake, we focus on a slightly simplified scenario, where all reactions have linear propensities (i.e., they are unimolecular in case of mass-action kinetics). As can be seen from Eq. (\ref{eq:finalpathmi}), the path mutual information involves inner expectations that are conditional on paths $X_0^t$, $Y_0^t$ and $\{X_0^t, Y_0^t\}$, as well as an outer expectation that integrates over different realizations of these paths. In our moment-approximation, the inner expectations are handled by finding approximate moments of the underlying filtering distribution, whereas the outer expectations are obtained by averaging over many (exact) stochastic simulations. To obtain analytical results, we further expand the $x\log x$ terms inside the outer expectations into a Taylor series. In particular, we obtain
\begin{equation}
\label{eq:taylor}
\lambda \log(\lambda) = \tilde{\lambda} \log (\tilde{\lambda}) + (\log (\tilde{\lambda})+1)(\lambda - \tilde{\lambda}) + \frac{(\lambda - \tilde{\lambda})^2}{2\tilde{\lambda}} + O(\lambda^3),
\end{equation}
where $\tilde{\lambda}$ is the point at which we perform the expansion. If we now consider $\lambda$ to be a random variable, set $\tilde{\lambda} = \mathbb{E}[\lambda]$ and take the expectation, we find
\begin{equation}
\label{eq:taylorwithexp}
\mathbb{E}[\lambda \log(\lambda)] \approx \mathbb{E}[\lambda] \log (\mathbb{E}[\lambda]) + (\log (\mathbb{E}[\lambda])+1)(\mathbb{E}[\lambda] - \mathbb{E}[\lambda]) + \frac{\mathbb{E}[(\lambda - \mathbb{E}[\lambda])^2]}{2\mathbb{E}[\lambda]},
\end{equation}
which simplifies to 
\begin{equation}
\mathbb{E}[\lambda \log (\lambda)] \approx \mathbb{E}[\lambda]\log(\mathbb{E}[\lambda])+\frac{\mathrm{Var}[\lambda]}{2\mathbb{E}[\lambda]}.
\end{equation}
Applying this approximation to (\ref{eq:finalpathmi}), we finally obtain
\begin{equation}
\label{eq:analyt}
\begin{split}
 I^{XY}_t \approx & \sum_{k \in R_{X}} \int_0^t \frac{\mathrm{Var}[\lambda^{X Y}_k(t)]-\mathrm{Var}[\lambda^{X}_k(t)]}{2\mathbb{E}[\lambda_k(Z(t))]} \mathrm{d}s + \sum_{k \in R_{Y}} \int_0^t \frac{\mathrm{Var}[\lambda^{X Y}_k(t)]-\mathrm{Var}[\lambda^{Y}_k(t)]}{2\mathbb{E}[\lambda_k(Z(t))]} \mathrm{d}s.
\end{split}
\end{equation}
This shows that up to second order, the path mutual information can be approximated essentially by the variances of the marginal propensities such as $\mathrm{Var}[\lambda_k^{XY}(t)] = \mathrm{Var}[\mathbb{E}[\lambda_k(Z(t))\mid X_0^t, Y_0^t]] $. The general idea is then to derive ordinary differential equations for these variances from the previously obtained conditional moment equations. In the following, we show how such equations can be derived. Concrete examples will be given later in Section 3.1.1.

For the sake of illustration, we consider the variance of the marginal propensity $\lambda_k^X(t) = \mathbb{E}[\lambda_k(Z(t)) \mid X_0^t]$. Since all reactions have linear rates, we have $\lambda_k (Z(t))=c_k \bar{Z}_{j(k)}(t)$ with $\bar{Z}_{j(k)}(t)$ as the copy number of the particular species in $\bar{\mathrm{Z}}$ that drives reaction $k$ and $c_k$ as a constant. As a consequence of the linear propensity functions, finding the marginal propensity $\lambda_k^X(t)$ is equivalent to finding the first conditional moment of $\bar{Z}_{j(k)}(t)$, which for compactness we denote by $M_k(t)=\mathbb{E}[\bar{Z}_{j(k)}(t) \mid X_0^t]$ in the following. 
Note that $M_k(t)$ is a functional of the path $X_0^t$ and as such is stochastic. The goal is now to find the variance of this stochastic process $\mathrm{Var}[M_k(t)]=\mathrm{Var}[\mathbb{E}[\bar{Z}_{j(k)}(t) \mid X_0^t]]$ from which we subsequently obtain $\mathrm{Var}[\lambda_k^X(t) ] = c_k^2 \mathrm{Var}[M_k(t)]$. 

The variance of $M_k(t)$ can be written as $\mathrm{Var}[M_k(t)] = \mathbb{E}[M_k(t)^2] - \mathbb{E}[M_k(t)]^2 =  \mathbb{E}[M_k(t)^2] - \mathbb{E}[\bar{Z}_{j(k)}(t)]^2$. Therefore, since the average $\mathbb{E}[\bar{Z}_{j(k)}(t)]$ is straightforward to obtain for linear networks, the calculation of the variance boils down to finding the second non-central moment of $M_k(t)$. To this end, we consider the differential equation of $M_k(t)$, which can be obtained by multiplying (\ref{eq:filtereq}) with $\bar{z}_{j(k)}$ and summing over all $\bar{z}$:
\begin{equation}
\begin{split}
\mathrm{d}M_k(t) &= \Big(D_k(t) - \sum_{i\in R_{X}} c_i \big(M_{i,k}(t)-M_i(t)M_k(t)\big)\Big)\mathrm{d}t + \sum_{i\in R_{X}}\frac{M_{i,k}(t)-M_i(t)M_k(t)}{M_i(t)}\mathrm{d}N_i(t) \\
&= \Big(D_k(t) - \sum_{i\in R_{X}} c_i C_{i,k}(t) \Big)\mathrm{d}t + \sum_{i\in R_{X}}\frac{ C_{i,k}(t)}{M_i(t)}\mathrm{d}N_i(t),
\end{split}
\label{eq:meanEq}
\end{equation}
where $D_k(t)= \sum_{z_j(k)} \mathcal{A}^{\bar{Z}\mid X}\pi^{X}(\bar{z},t)\cdot z_{j(k)}$ summarizes the fluxes originating from the forward operator $\mathcal{A}^{\bar{Z}\mid X}$ and where we have defined the second non-central conditional moment $M_{i,k}(t) := \mathrm{E}[Z_{j(i)}(t) Z_{j(k)}(t) \mid X_0^t]$ as well as the conditional covariance $C_{i,k}(t) :=  \mathrm{Cov}[Z_{j(i)}(t), Z_{j(k)}(t) \mid X_0^t] = M_{i,k}(t) - M_i(t)M_k(t)$. To obtain an equation for the second moment of $M_k(t)$, we first derive an equation for $M_k(t)^2$ using Ito's rule for counting processes \cite{lipster}, i.e., 
\begin{equation}
\mathrm{d}\bigl( M_k(t)^2\bigr)= 2M_k(t)\mathrm{d}M_k(t)+\sum_{i\in R_{X}} \frac{C_{i,k}(t)^2}{M_i(t)^2}\mathrm{d}N_i(t).
\end{equation}
Decomposing the reaction counters $\mathrm{d}N_i(t)$ into a predictable part and a martingale, i.e., $\mathrm{d}N_i(t)=c_i M_i(t) \mathrm{d}t + \mathrm{d}\tilde{N}_i(t)$, and inserting Eq. (\ref{eq:meanEq}) we further obtain
\begin{equation}
\begin{split}
\mathrm{d}\bigl( M_k(t)^2\bigr) & = 2 M_k(t) D_k(t) \mathrm{d}t - \sum_{i\in R_{X}} 2 c_i M_k(t)C_{i,k}(t) \mathrm{d}t  + \sum_{i\in R_{X}} 2 c_i M_k(t) C_{i,k}(t) \mathrm{d}t + \sum_{i\in R_{X}} 2 M_k(t) \frac{C_{i,k}(t)}{M_i(t)} \mathrm{d}\tilde{N}_i (t) \\ & +  \sum_{i\in R_{X}} c_i \frac{C_{i,k}(t)^2}{M_i(t)}\mathrm{d}t + \sum_{i\in R_{X}} \frac{C_{i,k}(t)^2}{M_i(t)^2}\mathrm{d}\tilde{N}_i(t) \\
&= 2 M_k(t) D_k(t) \mathrm{d}t + \sum_{i\in R_{X}} 2 M_k(t) \frac{C_{i,k}(t)}{M_i(t)} \mathrm{d}\tilde{N}_i (t) +  \sum_{i\in R_{X}} c_i \frac{C_{i,k}(t)^2}{M_i(t)}\mathrm{d}t + \sum_{i\in R_{X}} \frac{C_{i,k}(t)^2}{M_i(t)^2}\mathrm{d}\tilde{N}_i(t) .
\end{split}
\end{equation}
By taking the expectation, all terms involving the martingale increments $\mathrm{d}\tilde{N}_i(t)$ vanish, which after some simplifications leads to 
\begin{equation}
\label{eq:momsq}
\mathrm{d}\mathbb{E}[M_k(t)^2] = 2 \mathbb{E}[ M_k(t) D_k(t)] \mathrm{d}t + \sum_{i\in R_{X}}c_i\mathbb{E}\left[ \frac{C_{i,k}(t)^2}{M_i(t)} \right]\mathrm{d}t,
\end{equation}
Since the considered network architecture has linear propensities, the first term on the rhs of (\ref{eq:momsq}) will result in moments of at most order two, regardless of the specific form of $D_k(t)$. However, the second term involves a more complex expectation that is in general difficult to calculate. Interestingly, this expectation simplifies when the underlying filtering distribution is one-dimensional (i.e., $\bar{z}$ is a scalar that drives a single reaction $\bar{\mathrm{Z}}\rightarrow \bar{\mathrm{Z}} + \mathrm{X}$) and assumed to be a Gamma distribution. In this case, the sum in (\ref{eq:momsq}) involves only a single term of the form $\mathbb{E}[ C_{i,i}(t)^2 / M_i(t) ]$, whereas $C_{i, i}(t)$ is a conditional variance. Under the assumption that the filtering distribution is Gamma, it can then be shown that $\mathbb{E}[ C_{i,i}(t)^2 / M_i(t) ] = \mathbb{E}[C_{i,i}(t)]^2 / \mathbb{E}[M_i(t)]$. Inspired by this, and in line with the multivariate Gamma-closure (\ref{eq:GammaClosure}), we thus use the approximation $\mathbb{E}[ C_{i,k}(t)^2 / M_i(t) ] \approx \mathbb{E}[C_{i,k}(t)]^2 / \mathbb{E}[M_i(t)]$ to close Eq. (\ref{eq:momsq}). While this approximation is so far heuristic, it yielded very accurate results in our considered case studies.
\section{Case Studies}
\subsection{Case Study I}
\subsubsection{Path Mutual Information}
We consider a three-node feedforward network as presented in the main text. To obtain an expression for the path mutual information between species A and C, we realize that species C receives information about species $\mathrm{A}$ only through reaction $R_C=\{5\}$, which depends on intermediate species $\mathrm{B}$. Using Eq. (\ref{eq:finalpathmi}), the path mutual information takes the form   
\begin{equation}
 I_t^{AC} = \mathbb{E}\bigl[ \int_0^t k_5 \bigl(\mathbb{E}[B(s)\mid A_0^s ,C_0^s] \log (k_5 \mathbb{E}[B(s)\mid A_0^s , C_0^s]) - \mathbb{E}[B(s)\mid C_0^s]\log (k_5 \mathbb{E}[B(s)\mid C_0^s])\bigr) \mathrm{d}s\bigr]
\end{equation}
In this case, two filtering equations are required, one that describes $B(t)$ conditionally on $A_0^t$ and $C_0^t$ and one that describes $(A(t), B(t))$ conditionally on $C_0^t$. Based on Eq. (\ref{eq:filtereq}) these two filtering equations are given by 
\begin{equation}
\begin{split}
\label{eq:filter1}
\mathrm{d}\pi^{AC}(b,t) & =k_3 A(t) \left(\pi^{AC}(b-1,t)-\pi^{AC}(b,t)\right)\mathrm{d}t + k_4 \left((b+1)\pi^{AC}(b+1,t)-b\pi^{AC}(b,t)\right)\mathrm{d}t \\ 
& - k_5 \left(b-\mathbb{E}[B(t)\mid A_0^t , C_0^t]\right)\pi^{AC}(b,t)\mathrm{d}t + \frac{b-\mathbb{E}[B(t)\mid A_0^t , C_0^t]}{\mathbb{E}[B(t)\mid A_0^t , C_0^t]} \pi^{AC}(b,t)\mathrm{d}N_5(t)
\end{split}
\end{equation}
and 
\begin{equation}
\begin{split}
\mathrm{d}\pi^{C}(a,b,t) & = k_1 \left( \pi^{C}(a-1,b,t)-\pi^{C}(a,b,t)\right) \mathrm{d}t + k_2 \left( (a+1) \pi^{C}(a+1,b,t)-a \pi^{C}(a,b,t)\right) \mathrm{d}t \\
& + k_3 a \left(\pi^{C}(a,b-1,t)-\pi^{C}(a,b,t)\right)\mathrm{d}t + k_4 \left((b+1)\pi^{C}(a,b+1,t)-b\pi^{C}(a,b,t)\right)\mathrm{d}t \\ 
& - k_5 \left(b-\mathbb{E}[B(t)\mid C_0^t]\right)\pi^{C}(a,b,t)\mathrm{d}t  + \frac{b-\mathbb{E}[B(t)\mid C_0^t]}{ \mathbb{E}[B(t)\mid C_0^t]} \pi^{C}(a,b,t)\mathrm{d}N_5(t)
\end{split}
\end{equation}
which in principle can be solved numerically on a finite grid using the quasi-exact method. Moreover, differential equations can be obtained from these equations for the conditional means $\mathbb{E}[B(t)\mid A_0^t ,C_0^t]$ and $\mathbb{E}[B(t)\mid C_0^t]$ which will, however, depend on moments of higher order. In case of Eq. (\ref{eq:filter1}), for instance, we obtain for the first two conditional moments
\begin{equation}
\begin{split}
\mathrm{d}\mathbb{E}[B(t)\mid A_0^t , C_0^t] & = k_3 A(t) \mathrm{d}t - k_4 \mathbb{E}[B(t)\mid A_0^t , C_0^t] \mathrm{d}t - k_5 \mathbb{E}[B(t)^2\mid A_0^t , C_0^t] \mathrm{d}t + k_5 \mathbb{E}[B(t)\mid A_0^t ,C_0^t]^2 \mathrm{d}t \\ & + \frac{\mathbb{E}[B(t)^2\mid A_0^t ,C_0^t]-\mathbb{E}[B(t)\mid A_0^t ,C_0^t]^2}{\mathbb{E}[B(t)\mid A_0^t ,C_0^t]}\mathrm{d}N_5(t)\\ 
 \mathrm{d}\mathbb{E}[B(t)^2\mid A_0^t ,C_0^t] & = 2k_3 A(t) \mathbb{E}[B(t)\mid A_0^t ,C_0^t]\mathrm{d}t+k_3 A(t) \mathrm{d}t - 2k_4\mathbb{E}[B(t)^2\mid A_0^t ,C_0^t]\mathrm{d}t + k_4 \mathbb{E}[B(t)\mid A_0^t ,C_0^t]\mathrm{d}t \\ & 
- k_5 \mathbb{E}[B(t)^3 \mid A_0^t , C_0^t]\mathrm{d}t + k_5 \mathbb{E}[B(t)^2\mid A_0^t , C_0^t] \mathbb{E}[B(t)\mid A_0^t , C_0^t]\mathrm{d}t \\ &
+ \frac{\mathbb{E}[B(t)^3 \mid A_0^t , C_0^t]-\mathbb{E}[B(t)^2\mid A_0^t , C_0^t] \mathbb{E}[B(t)\mid A_0^t , C_0^t]}{\mathbb{E}[B(t)\mid A_0^t , C_0^t]}\mathrm{d}N_5(t).
\end{split} 
\end{equation}
This implies that one would need an infinite amount of differential equations in order to calculate the moment dynamics. Applying a third-order Gamma-closure allows us replace the third conditional moment $\mathbb{E}[B(t)^3 \mid A_0^t , C_0^t]$ by moments of order one and two using (\ref{eq:GammaClosure}). Then, the closed equation for the second moment becomes
\begin{equation}
\begin{split}
 \mathrm{d}\mathbb{E}[B(t)^2\mid A_0^t ,C_0^t] & = 2k_3 A(t) \mathbb{E}[B(t)\mid A_0^t ,C_0^t]\mathrm{d}t+k_3 A(t) \mathrm{d}t - 2k_4\mathbb{E}[B(t)^2\mid A_0^t ,C_0^t]\mathrm{d}t + k_4 \mathbb{E}[B(t)\mid A_0^t ,C_0^t]\mathrm{d}t \\ & - 2k_5 \frac{\mathbb{E}[B(t)^2\mid A_0^t ,C_0^t]^2}{\mathbb{E}[B(t)\mid A_0^t ,C_0^t]}\mathrm{d}t +2k_5 \mathbb{E}[B(t)\mid A_0^t ,C_0^t]\mathbb{E}[B(t)^2\mid A_0^t ,C_0^t]\mathrm{d}t \\ & + 
  2\frac{\mathbb{E}[B(t)^2\mid A_0^t ,C_0^t]^2}{\mathbb{E}[B(t)\mid A_0^t ,C_0^t]^2}\mathrm{d}N_5(t)-2\mathbb{E}[B(t)^2\mid A_0^t ,C_0^t]\mathrm{d}N_5(t).
\end{split} 
\end{equation} 
In the same way, we can get equations for the set of differential equations conditionally on $C_0^t$ to evaluate $\mathbb{E}[B(t)\mid C_0^t]$.

In this example, we also made use of the analytical approximation of the path mutual information (i.e., Eq. (\ref{eq:analyt})) by which we obtain 
\begin{equation}
\label{eq:ACAnalytical}
I_t^{AC} \approx\int_0^t \frac{k_5}{2}\frac{\mathrm{Var}[\mathbb{E}[B(s)\mid A_0^s ,C_0^s]]-\mathrm{Var}[\mathbb{E}[B(s)\mid C_0^s]]}{\mathbb{E}[B(s)]}\mathrm{d}s.
\end{equation}
In order to demonstrate how to obtain the variances of the respective conditional means, we derive $\mathrm{Var}[\mathbb{E}[B(s)\mid A_0^s ,C_0^s]]$ as discussed in section 2.2. We start with defining the differential
\begin{equation*}
\begin{split}
\mathrm{d}\mathrm{Var}[\mathbb{E}[B(t)\mid A_0^t ,C_0^t]]&=\mathrm{d}\mathbb{E}[\mathbb{E}[B(t)\mid A_0^t ,C_0^t]^2]-\mathrm{d}\mathbb{E}[\mathbb{E}[B(t)\mid A_0^t ,C_0^t]]^2\\ & =\mathrm{d}\mathbb{E}[\mathbb{E}[B(t)\mid A_0^t ,C_0^t]^2]-2\mathbb{E}[B(t)]\mathrm{d}\mathbb{E}[B(t)].
\end{split}
\end{equation*} 
The dynamics of $\mathbb{E}[B(t)]$ can be derived from a conventional master equation and is given by 
\begin{equation}
\label{eq:meb}
\mathrm{d}\mathbb{E}[B(t)]=\bigl(k_3\mathbb{E}[A(t)]-k_4\mathbb{E}[B(t)]\bigr)\mathrm{d}t.
\end{equation}
Using Eq. (\ref{eq:momsq}) and realizing that 
\begin{equation}
\begin{split}
\mathbb{E}[A(t)\mathbb{E}[B(t)\mid A_0^t , C_0^t]] & = \mathbb{E}[\mathbb{E}[A(t) B(t)\mid A_0^t , C_0^t]] = \mathbb{E}[A(t)B(t)],
\end{split}
\end{equation}
we get an approximate expression for the differential of the squared moment
\begin{equation}
\label{eq:sqdiff}
\frac{\mathrm{d}}{\mathrm{d}t}\mathbb{E}[\mathbb{E}[B(t)\mid A_0^t ,C_0^t]^2]=2k_3  \mathbb{E}[A(t) B(t)]-2k_4\mathbb{E}[\mathbb{E}[B(t)\mid A_0^t , C_0^t]^2]+k_5\frac{\mathbb{E}[\mathrm{Var}[B(t)\mid A_0^t ,C_0^t]]^2}{\mathbb{E}[B(t)]},
\end{equation}
where we have used the approximation $\mathbb{E}[ \mathrm{Var}[B(t)\mid A_0^t ,C_0^t]^2/\mathbb{E}[B(t) \mid A_0^t ,C_0^t] ] \approx \mathbb{E}[\mathrm{Var}[B(t)\mid A_0^t ,C_0^t]]^2 / \mathbb{E}[B(t)]$. Now, we can combine Eqs. (\ref{eq:meb}) and (\ref{eq:sqdiff}) and by subtracting $2\mathbb{E}[B(t)\mid A_0^t ,C_0^t]\mathrm{d}\mathbb{E}[B(t)\mid A_0^t ,C_0^t]=2\mathbb{E}[B(t)]\mathrm{d}\mathbb{E}[B(t)]$, find that the requested variance satisfies the differential equation 
\begin{equation}
\begin{split}
\frac{\mathrm{d}}{\mathrm{d}t}\mathrm{Var}[\mathbb{E}[B(t)\mid A_0^t ,C_0^t]]& =\Bigl(2k_3 \mathrm{Cov}[A(t),B(t)]-2k_4\mathrm{Var}[\mathbb{E}[B(t)\mid A_0^t ,C_0^t]] +k_5 \frac{\mathbb{E}[\mathrm{Var}[B(t)\mid A_0^t ,C_0^t]]^2}{\mathbb{E}[B(t)]}\Bigr).
\end{split}
\label{eq:VarExp}
\end{equation}
Eq. (\ref{eq:VarExp}) involves additional moments, for which equations can be obtained in a similar way. In total, this leads to a system of differential equations
\begin{equation}
\begin{split}
\frac{\mathrm{d}}{\mathrm{d}t}\mathrm{Var}[\mathbb{E}[B(t)\mid A_0^t ,C_0^t]] &=\Bigl(2k_3 \mathrm{Cov}[A(t),B(t)]-2k_4\mathrm{Var}[\mathbb{E}[B(t)\mid A_0^t ,C_0^t]] +k_5 \frac{\mathbb{E}[\mathrm{Var}[B(t)\mid A_0^t ,C_0^t]]^2}{\mathbb{E}[B(t)]}\Bigr) \\
\frac{\mathrm{d}}{\mathrm{d}t}\mathbb{E}[A(t)]&=\bigl(k_1-k_2\mathbb{E}[A(t)]\bigr)\\
\frac{\mathrm{d}}{\mathrm{d}t}\mathbb{E}[B(t)]&=\bigl(k_3\mathbb{E}[A(t)]-k_4\mathbb{E}[B(t)]\bigr) \\
\frac{\mathrm{d}}{\mathrm{d}t}\mathbb{E}[\mathrm{Var}[B(t)\mid A_0^t ,C_0^t]]&=\Bigl(k_3\mathbb{E}[A(t)]+k_4\mathbb{E}[B(t)]-2k_4\mathbb{E}[\mathrm{Var}[B(t)\mid A_0^t ,C_0^t]]-k_5 \frac{\mathbb{E}[\mathrm{Var}[B(t)\mid A_0^t ,C_0^t]]^2}{\mathbb{E}[B(t)]}\Bigr)\\
\frac{\mathrm{d}}{\mathrm{d}t}\mathrm{Cov}[A(t),B(t)]&=\bigl(-(k_2+k_4)\mathrm{Cov}[A(t),B(t)]+k_3\mathrm{Var}[A(t)]\bigr)\\
\frac{\mathrm{d}}{\mathrm{d}t}\mathrm{Var}[A(t)]&=\bigl(k_1+k_2\mathbb{E}[A(t)]-2k_2 \mathrm{Var}[A(t)]\bigr),
\end{split}
\end{equation}
whose steady state can be found by setting the lhs to zero. The same procedure can be followed for calculating the variance $\mathrm{Var}[\mathbb{E}[B(t)\mid C_0^t]]$, which in combination with $\mathrm{Var}[\mathbb{E}[B(t)\mid A_0^t ,C_0^t]]$ allows us to approximate the path mutual information via Eq. (\ref{eq:ACAnalytical}) at steady state. Note that for stationary systems, the path mutual information rate can be obtained simply by dropping the time integral in (\ref{eq:ACAnalytical}). Doing so, and considering the limit $v_c \rightarrow \infty$ (or equivalently $k_5$) leads to the steady state mutual information rate shown in Eq. (9) in the main text.

\begin{figure}
\centering
\includegraphics[width=0.99\textwidth]{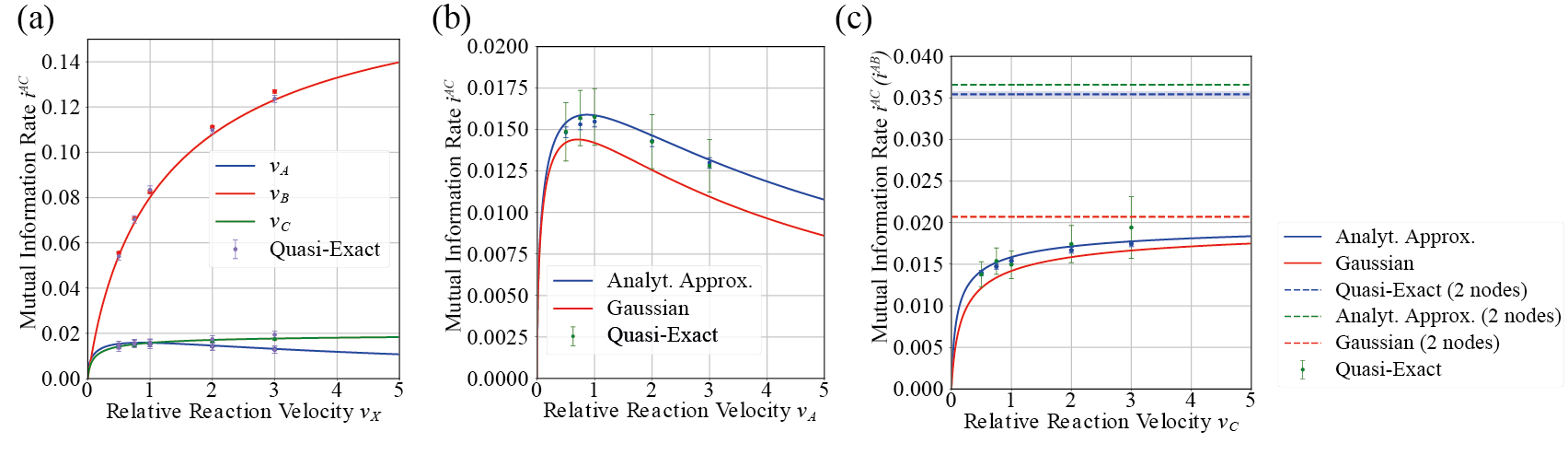}
\caption{(a) Analytical approximation of the path mutual information rate $i^{AC}$ as a function of the relative reaction velocity $v_X$, $X\in [A,B,C]$ and comparison to the quasi-exact solution (purple points). (b) Path mutual information rate $i^{AC}$ as a function of $v_A$. The quasi-exact solution is given by the green points. (c) Path mutual information rate $i^{AC}$ as a function of $v_C$ for the network consisting of three nodes and the one of two nodes.  The simulations are performed with the same set of parameters as for Figure 2, main text. The quasi-exact calculation is performed with a sample size of n=600. The points colored corresponding to the analytical approximated curves correspond to the respective solution obtained via moment closure approximation with a sample size of $n=10000$.}
\label{fig:fig2-appendix}
\end{figure}

In Fig.~2 in the main text, we compared the analytical approximation with the moment-approximation method and found very good agreement. However, since also the moment-approximation method is approximate, we carried out a similar comparison with the quasi-exact method (Fig. ~\ref{fig:fig2-appendix}). Also in this case, we found very good agreement between the two approaches. Note that due to the computational complexity of the quasi-exact method, this analysis was restricted to a smaller parameter region than the one considered in Fig.~2 in the main text. 

In the main text, we compared the mutual information rate between species $\mathrm{A}$ and $\mathrm{C}$ with the mutual information rate between $\mathrm{A}$ and $\mathrm{B}$. According to Eq. (\ref{eq:finalpathmi}), the latter is given by
\begin{equation}
\label{eq:MIAB}
 I_t^{AB} = \mathbb{E}\bigl[ \int_0^t k_3 (A(s) \log (k_3 A(s)) - \mathbb{E}[A(s)\mid B_0^s]\log (k_3 \mathbb{E}[A(s)\mid B_0^s])) \mathrm{d}s\bigr],
\end{equation}
where information transfer is mediated through reaction $R_B=\{3\}$.
%In this case, all the information between those species is transmitted via reaction. 
Note that the conditional expectations vanish in the first term on the rhs of (\ref{eq:MIAB}), because conditioning on $\{A_0^t, B_0^t\}$ provides complete knowledge about the history of the network (i.e., $\mathbb{E}[A(t) \mid A_0^t, B_0^t] = A(t)$). The analytical approximation of the path mutual information then becomes
\begin{equation}
\label{eq:analytAB}
I_t^{AB} \approx\int_0^t \frac{k_3}{2}\frac{\mathrm{Var}[A(s)]-\mathrm{Var}[\mathbb{E}[A(s)\mid B_0^s]]}{\mathbb{E}[A(s)]}\mathrm{d}s.
\end{equation}
At stationarity, the path mutual information rate is then given by the integrand of Eq. (\ref{eq:analytAB}). Moreover, by the law of total variance, it holds that $\mathrm{Var}[A(s)]-\mathrm{Var}[\mathbb{E}[A(s)\mid B_0^s]] = \mathbb{E}[\mathrm{Var}[A(t)\mid B^t_0]]$, such that we obtain
\begin{equation}
\label{eq:lotv}
\lim_{t\rightarrow \infty} i_t^{AB} = i^{AB}=\frac{k_3}{2}\frac{\mathbb{E}[\mathrm{Var}[A(t)\mid B^t_0]]}{\mathbb{E}[A(t)]}.
\end{equation}
The steady state of $\mathbb{E}[A(t)]$ can be obtained directly from the underlying master equation and is given by $\lim_{t\rightarrow \infty} \mathbb{E}[A(t)] = k_1/k_2$. Therefore, the only term that needs to be determined is the expected conditional variance $\mathbb{E}[\mathrm{Var}[A(t)\mid B^t_0]]$ satisfies the differential equation
\begin{equation}
\frac{\mathrm{d}}{\mathrm{d}t}\mathbb{E}[\mathrm{Var}[A(t)\mid B^t_0]]=k_1+k_2\mathbb{E}[A(t)]-2k_2\mathbb{E}[\mathrm{Var}[A(t)\mid B^t_0]]-k_3 \mathbb{E}\Bigl[\frac{\mathrm{Var}[A(t)\mid B^t_0]^2}{\mathbb{E}[A(t) \mid B_0^t]}\Bigr],
\end{equation}
which can be derived using Ito's rule for counting processes as demonstrated before. Making use of the approximation $\mathbb{E}[\mathrm{Var}[A(t)\mid B^t_0]^2/\mathrm{E}[A(t) \mid B_0^t] \approx \mathbb{E}[\mathrm{Var}[A(t)\mid B^t_0]]^2/\mathbb{E}[A(t)]$, solving for the stationary solution and inserting it into (\ref{eq:lotv}) finally yields
\begin{equation}
i^{AB}=-\frac{k_2}{2}+\frac{1}{2}\sqrt{k_2(k_2+2k_3)}.
\end{equation}

\subsubsection{Gaussian Approximations}
For comparison, we calculated stationary mutual information rates using Gaussian theory as proposed by Tostevin et al. \cite{PhysRevLett.102.218101}. To this end, we employ the relationship
\begin{equation}
\label{eq:tw}
i_G^{XY}=-\frac{1}{4\pi}\int_{-\infty}^{\infty} \log \left[ 1 - \frac{|S_{XY}(\omega)|^2}{S_{XX}(\omega)S_{YY}(\omega)}\right] \mathrm{d}\omega
\end{equation}
where $S_{XY}(\omega )$ is the cross power spectral density of signals $X(t)$ and $Y(t)$ and $S_{XX}(\omega)$ and $S_{YY}(\omega)$ are the power spectral densities of $X(t)$ and $Y(t)$, respectively. Since all propensities in the considered two- and three-node networks are linear, $S_{XY}(\omega )$, $S_{XX}(\omega)$ and $S_{YY}(\omega)$ can be calculated in closed form, by first deriving the respective cross- and auto-covariance functions and calculating their Fourier transforms. For the considered two-node network, we obtain
\begin{equation}
\label{eq:covarAB}
\begin{split}
\frac{\mathrm{d}\mathrm{Cov}[A(0),A(t)]}{\mathrm{d}t} & = -k_2 \mathrm{Cov}[A(0),A(t)] \\
\frac{\mathrm{d}\mathrm{Cov}[A(0),B(t)]}{\mathrm{d}t} & =k_3\mathrm{Cov}[A(0),A(t)]-k_4\mathrm{Cov}[A(0),B(t)] \\
\frac{\mathrm{d}\mathrm{Cov}[B(0),B(t)]}{\mathrm{d}t} & = k_3\mathrm{Cov}[B(0),A(t)] -k_4 \mathrm{Cov}[B(0),B(t)]\\
\frac{\mathrm{d}\mathrm{Cov}[B(0),A(t)]}{\mathrm{d}t} & = - k_2\mathrm{Cov}[B(0),A(t)].
\end{split}
\end{equation}
For the three-node network, we obtain
\begin{equation}
\label{eq:covarAC}
\begin{split}
\frac{\mathrm{d}\mathrm{Cov}[A(0),A(t)]}{\mathrm{d}t} & = -k_2 \mathrm{Cov}[A(0),A(t)]\\
\frac{\mathrm{d}\mathrm{Cov}[A(0),B(t)]}{\mathrm{d}t} & =k_3\mathrm{Cov}[A(0),A(t)]-k_4\mathrm{Cov}[A(0),B(t)] \\
\frac{\mathrm{d}\mathrm{Cov}[A(0),C(t)]}{\mathrm{d}t} & =k_5\mathrm{Cov}[A(0),B(t)]-k_6\mathrm{Cov}[A(0),C(t)] \\
\frac{\mathrm{d}\mathrm{Cov}[C(0),A(t)]}{\mathrm{d}t} &=  - k_2\mathrm{Cov}[C(0),A(t)] \\ 
\frac{\mathrm{d}\mathrm{Cov}[C(0),A(t)]}{\mathrm{d}t} &= k_3 \mathrm{Cov}[C(0),A(t)] - k_4 \mathrm{Cov}[C(0),B(t)] \\
\frac{\mathrm{d}\mathrm{Cov}[C(0),C(t)]}{\mathrm{d}t} & =k_5\mathrm{Cov}[C(0),B(t)]-k_6\mathrm{Cov}[C(0),C(t)].
\end{split}
\end{equation}
Eqs. (\ref{eq:covarAB}) and (\ref{eq:covarAC}) define systems of ordinary differential equations, which can be Fourier-transformed directly, giving rise to linear systems of algebraic equations. Solving these equations yields $S_{AB}(\omega ), \: S_{AA}(\omega )$ and $S_{BB}(\omega )$ in case of the two-node network and $S_{AC}(\omega ), \: S_{AA}(\omega )$ in case of the three-node network $S_{CC}(\omega )$, respectively. In that way, the fraction $F_{XY}(\omega)=|S_{XY}(\omega)|^2/(S_{XX}(\omega)S_{YY}(\omega))$ inside (\ref{eq:tw}) bcomes
\begin{equation}
\label{eq:tw2}
F_{AB}(\omega)=\frac{k_2 k_3}{\omega ^2+k_2 (k_2 + k_3)},
\end{equation}
in case of the two-node network, and
\begin{equation}
\label{eq:tw3}
F_{AC}(\omega)=\frac{k_2 k_3 k_4 k_5}{k_2 k_4 (k_3 k_5 + k_2 (k_4 + k_5))+ (k_2^2+k_4(k_4+k_5))\omega ^2+\omega ^4},
\end{equation}
in case of the three-node network.
In the former case, the remaining integral in (\ref{eq:tw}) can be solved analytically, leading to
\begin{equation}
\label{eq:twr}
i^{AB}_G=-\frac{k_2}{2}+\frac{1}{2}\sqrt{k_2(k_2+k_3)}.
\end{equation} 
Taking the limit $k_5\rightarrow \infty$ in (\ref{eq:tw3}) directly leads to (\ref{eq:tw2}), showing that $\lim_{k_5\rightarrow \infty} i_G^{AC} = i_G^{AB}$. In other words, the three-node network reduces to the simpler two-node network in terms of information transfer as $k_5\rightarrow \infty$. Note that in case of the three-node network, inserting (\ref{eq:tw3}) into (\ref{eq:tw}) led to an integral, which could not be solved analytically, which is why an explicit expression of $i_G^{AC}$ could not be provided. For the results shown in Fig.~2e in the main text, this integral was solved numerically.

\subsection{Case Study II}
The second system considered in the main text is a bistable switch. The system is similar to three-node feedforward network, but contains a positive feedback between species $\mathrm{C}$ and $\mathrm{A}$. This feedback is incorporated by letting $k_1$ depend on the abundance of $\mathrm{C}$. In particular, we chose a Hill-type propensity of the form $k_1(C(t))=\mu C(t)^{n_H}/(K^{n_H}+C(t)^{n_H})+\epsilon$ with Hill coefficient $n_H=3$ and $\mu$ and $\epsilon$ as positive constants. Using Eq.~(\ref{eq:finalpathmi}), the path mutual information between A and C takes the form
\begin{equation}
\begin{split}
I_t^{AC} & = \mathbb{E}\bigl[ \int_0^t k_1(C(s))\log(k_1(C(s))) - \mathbb{E}[k_1(C(s))\mid A_0^s]\log(\mathbb{E}[k_1(C(s))\mid A_0^s]) \mathrm{d}s \\ & + k_5 \int_0^t \mathbb{E}[B(s)\mid A_0^s ,C_0^s]\log(\mathbb{E}[B(s)\mid A_0^s ,C_0^s])-\mathbb{E}[B(s)\mid C_0^s]\log(\mathbb{E}[B(s)\mid C_0^s])\bigr]\mathrm{d}s
\end{split}
\end{equation}
In contrast to the previous cases, information transmission is now bidirectional mediated through reactions $R_A=\{1\}$ and $R_C=\{5\}$. In this case, we require three filtering equations to calculate the path mutual information:
\begin{equation}
\begin{split}
\mathrm{d}\pi^{AC}(b,t) & =k_3 A(t) (\pi^{AC}(b-1,t)-\pi^{AC}(b,t))\mathrm{d}t + k_4((b+1)\pi^{AC}(b+1,t)-b\pi^{AC}(r,t))\mathrm{d}t \\ 
& - k_5(b-\mathbb{E}[B(t)\mid A_0^t ,C_0^t])\pi^{AC}(b,t)\mathrm{d}t + \frac{b-\mathbb{E}[B(t)\mid A_0^t ,C_0^t]}{\mathbb{E}[B(t)\mid A_0^t ,C_0^t]}\pi^{AC}(b,t)\mathrm{d}N_5 (t)
\end{split}
\end{equation}
\begin{equation}
\label{eq:filtA}
\begin{split}
\mathrm{d}\pi^{A}(b,c,t) & =  k_3 A(t) (\pi^{A}(b-1,c,t)-\pi^{A}(b,c,t))\mathrm{d}t + k_4((b+1)\pi^{A}(b+1,c,t)-b\pi^{A}(b,c,t))\mathrm{d}t \\ 
& + k_5b(\pi^{A}(b,c-1,t)-\pi^{A}(b,c,t))\mathrm{d}t  + k_6 ( (c+1)\pi^{A}(b,c+1,t)-c\pi^{A}(b,c,t)) \mathrm{d}t \\
& - (k_1(c)-\mathbb{E}[k_1(C(t))\mid A_0^t])\pi^{A}(b,c,t)\mathrm{d}t + \frac{k_1(c)-\mathbb{E}[k_1(C(t))\mid A_0^t]}{\mathbb{E}[k_1(C(t))\mid A_0^t]}\pi^{A}(b,c,t)\mathrm{d}N_1 (t)
\end{split}
\end{equation}
\begin{equation}
\begin{split}
\mathrm{d}\pi^{C}(a,b,t)& =k_1(C(t)) (\pi^{C}(a-1,b,t)-\pi^{C}(a,b,t))\mathrm{d}t + k_2 ((a+1)\pi^{C}(a+1,b,t)-a\pi^{C}(a,b,t))\mathrm{d}t\\
&+ k_3 a (\pi^{C}(a,b-1,t)-\pi^{C}(a,b,t))\mathrm{d}t+ k_4 ((b+1)\pi^{C}(a,b+1,t)-b\pi^{C}(a,b,t))\mathrm{d}t\\
&- k_5(b-\mathbb{E}[B(t)\mid C_0^t])\pi^{C}(a,b,t)\mathrm{d}t+\frac{b-\mathbb{E}[B(t)\mid C_0^t]}{\mathbb{E}[B(t)\mid C_0^t]}\pi^{C}(a,b,t)\mathrm{d}N_5 (t).
\end{split}
\label{eq:filter3}
\end{equation}
To calculate the path mutual information between species $\mathrm{A}$ and $\mathrm{C}$ for various parameter regimes, we made use of the moment-approximation method. However, since  $k_1(C(t))$ is not of polynomial form, this method is not directly applicable to the third filtering equation Eq. (\ref{eq:filtA}), because the rhs of the resulting moment dynamics would involve expectations of rational functions (and thus, no longer depend on moments only). To address this issue, we linearize $k_1(C(t))$ around the conditional expectation $\mathbb{E}[C(t)\mid A_0^t]$, which yields
\begin{equation}
\begin{split}
&k_1(C(t))\approx \\
&\quad \frac{\mu\mathbb{E}[C(t)\mid A_0^t]^3 }{K^3+\mathbb{E}[C(t)\mid A_0^t]^3}+\epsilon  + \Bigl( \frac{3\mu\mathbb{E}[C(t)\mid A_0^t]^2}{K^3+\mathbb{E}[C(t)\mid A_0^t]^3}-\frac{3\mu \mathbb{E}[C(t)\mid A_0^t]^5}{(K^3+\mathbb{E}[C(t)\mid A_0^t]^3)^2}\Bigr) \bigl(C(t)-\mathbb{E}[C(t)\mid A_0^t]\bigr).
\end{split}
\end{equation}

Moments can then be derived from (\ref{eq:filtA}) as described before. This approximation is expected to be accurate when the filtering distribution shows little variation in $C(t)$ around its mean $\mathbb{E}[C(t)\mid A_0^t]$, which should be the case when reaction $R_A=\{1\}$ fires frequently (i.e., observing this reaction frequently, reveals more information about about $C(t)$). We remark that this approximation needs to be performed only for (\ref{eq:filtA}) since the other two equations are conditioned on $C_0^t$, such that no averaging over $C(t)$ has to be performed (which would otherwise give rise to expectations over rational functions). 
\begin{figure}
\centering
\includegraphics[width=0.3\textwidth]{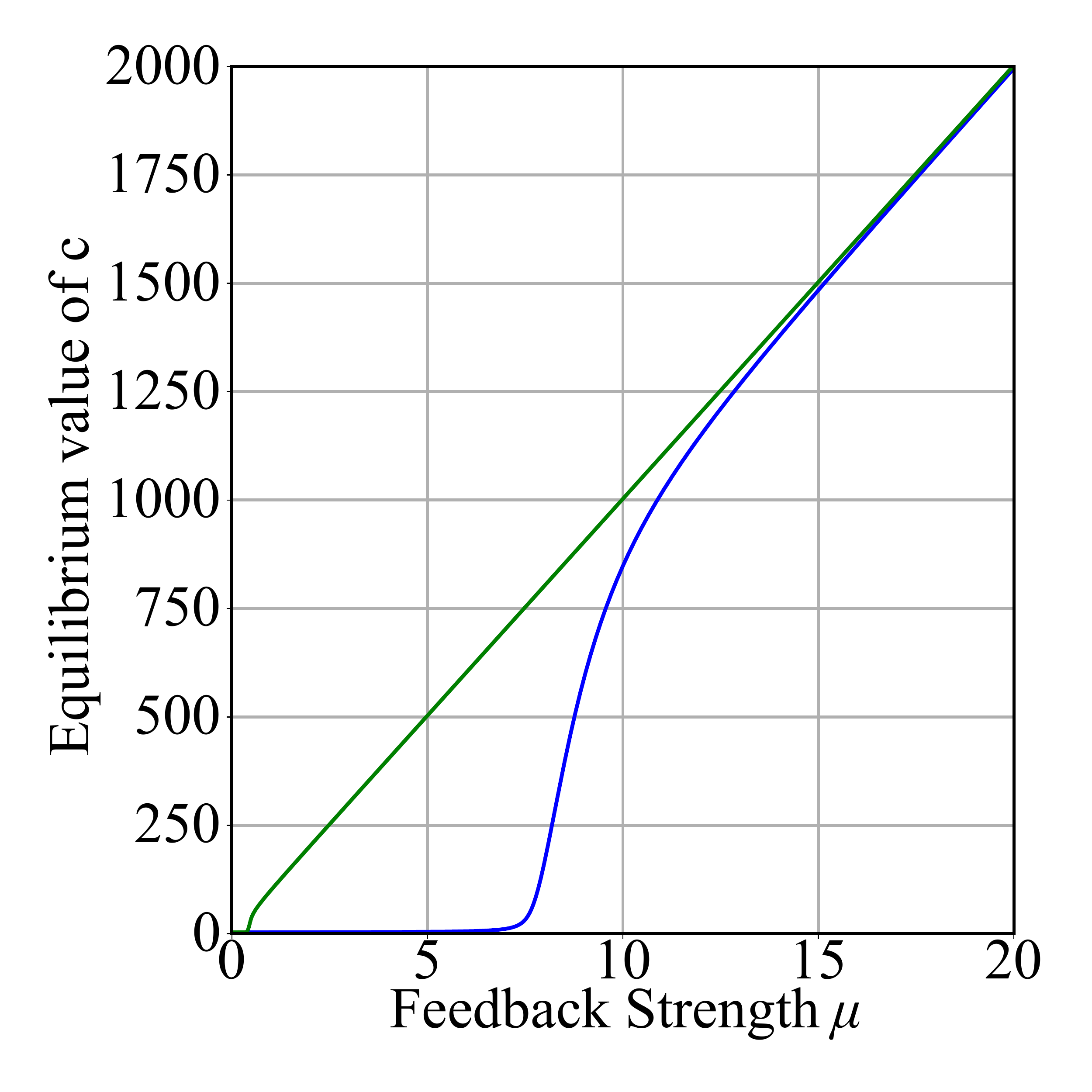}
\caption{Equilibrium points of the positive-feedback system as a function of $\mu$ predicted by mean-field theory. The bifurcation point is at $\mu\approx 0.35$.}
\label{fig:bif}
\end{figure}
In order to simplify the identification of parameter regimes where this system exhibits two meta-stable states, we analyzed the corresponding macroscopic mean-field equations given by 
\begin{equation}
\begin{split}
 \frac{\mathrm{d}}{\mathrm{d}t}a(t) & = \frac{\mu c(t)^3}{K^3+c(t)^3}+\epsilon - k_2 a(t) \\ 
 \frac{\mathrm{d}}{\mathrm{d}t} b(t) & = k_3 a(t) - k_4 b(t) \\
 \frac{\mathrm{d}}{\mathrm{d}t} c(t) & = k_5 b(t) - k_6 c(t) 
 \end{split}
\end{equation}
Fig.~\ref{fig:bif} shows the equilibrium points of this system as a function of the feedback strength $\mu$ for the parameters $\epsilon=0.03, K=30, k_2= 0.1, k_3 = 1, k_4=0.1, k_5=0.1, k_6=0.1$. This shows that there are two (stable) equilibrium points once $\mu$ exceeds a threshold ($\mu\approx0.35$), which subsequently approach each other when $\mu$ increases further.

\newpage

%\bibliographystyle{ieeetr}  % to be commented out when using main.bbl
%\bibliography{aipsamp} % to be commented out when using main.bbl